\documentclass[%
 reprint,
 amsmath,amssymb,
 aps,
]{revtex4-2}

\usepackage{graphicx}% Include figure files
\usepackage{dcolumn}% Align table columns on decimal point
\usepackage{bm}% bold math
\usepackage{epstopdf}
\usepackage{hyperref}% add hypertext capabilities
\begin{document}

\preprint{APS/123-QED}

\title{Fault-tolerant measurement-device-independent quantum key distribution \\with noisy non-Gaussian error correction}

\author{Zhiyue Zuo}
\affiliation{School of Automation, Central South University, Changsha 410083, China}
\author{Stefano Pirandola}
\email{stefano.pirandola@york.ac.uk}
\affiliation{Department of Computer Science, University of York, York YO10 5GH, United Kingdom}

\date{\today}% It is always \today, today,
             
\begin{abstract}
It is well known that the repeater node is an essential ingredient for the future global quantum network, which will enable high-rate private communication and entanglement distribution over very long distances.
The near-term repeater architecture uses the measurement-based node that operate without both entanglement and quantum memory, which is the main idea of the measurement-device-independent quantum key distribution (MDI-QKD) protocol.
The MDI-QKD protocol removes the trust condition from the inter repeaters, while its continuous variable (CV) version, when proposed, benefited from its deterministic nature, compatible with the classical devices, and shows a high rate for the short-range local area network (LAN).
Whilst the theoretical backbone of CV-MDI-QKD protocol is well established, its secure transmission range is yet limited for practical LAN.
In this study, we propose an enhanced scheme for the asymmetric CV-MDI-QKD protocol by using Gottesman-Kitaev-Preskill (GKP) oscillators-to-oscillators codes, where both loss error and operation error are suppressed to below the break-even point, without any delays caused by classical heralding signals.
In particular, the proposed scheme, which correlates the noises of the data and ancilla via a pair of symplectic transforms, extracts the error syndromes from stabilizer measurements on the ancilla mode and informs the data mode for a corrective displacement.
Numerical analysis shows the composable finite-size security of the protocol under the collective Gaussian attack, encompassing noiseless and noisy GKP states, with both wired (i.e., fiber-based) and wireless (i.e., free-space) configurations. 
In addition, we demonstrate that the residual errors of the GKP code can be further reduced by the concatenation method but has a trade-off between the layers number and the finite GKP squeezing.
\end{abstract}

\maketitle

%\tableofcontents

\section{Introduction}

% 引入到MDI（从量子网络出发）

%The Internet, designed to facilitate global communication and distributed information processing, has become an invaluable socioeconomic fixture as the number of devices increases exponentially over time.
In the long term, it is widely recognized that quantum internet will complement the traditional internet, which will be provably secure and could provide exponentially more computational power and sensing capability for specific tasks~\cite{kimble2008quantum,pirandola2016physics,wehner2018quantum}.
In detail, the quantum internet is going to be a hybrid ecosystem where different approaches and materials need to be suitably interfaced and simultaneously optimal, such as quantum transduction, quantum storage, and quantum communication.

Generally, the security layers are at the basis of quantum internet~\cite{azuma2023quantum}.
Recently, the integration of quantum key distribution (QKD) is gradually solving this problem~\cite{pirandola2020advances,GisinRMP}, realizing the required migration to near-term quantum-safe communication networks.
However, even though QKD creates information-theoretic security, the inherent fragility of quantum information makes achieving high rates over long distances much more challenging.
Pirandola \textit{et al.} demonstrated that $- {\log _2}\left( {1 - \tau } \right)$, where $ \tau $ is the channel’s transmittance, is the maximum fundamental rate, in bits, at which two distant parties can distribute quantum bits, entanglement bits, or secret bits~\cite{pirandola2017fundamental}. 
This is known as the Pirandola-Laurenza-Ottaviani-Banchi (PLOB) bound, which holds for any point-to-point protocol of quantum communication and has recently been extended to end-to-end variants for quantum networks~\cite{pirandola2019end,harney2022analytical,harney2022end}.
To beat the PLOB bound, it is necessary to insert in-the-middle quantum relay.
To date, the most advanced relay is the entanglement-based node, which is generally-untrusted and has the capability to perform entanglement swapping, quantum store, and entanglement distillation~\cite{pirandola2022architectures}.
Unfortunately, the realization of such a node remains challenging due to factors such as quantum memory technology, the limited brightness of entanglement sources, and technological restrictions associated with distillation protocols~\cite{azuma2023quantum}.

Besides the entanglement-based node, a promising solution with current technology is the use of a measurement-based (MB) node, which is also untrusted but receives-only.
In detail, Alice and Bob send their quantum systems to the MB node, which measures them and broadcasts the classical outcome to create shared randomness.
A typical application of such a node is the measurement-device-independent (MDI) QKD protocol, where the node exploits a Bell detection~\cite{BP_MDI,lo2012measurement}. 
However, neither the discrete-variable (DV) nor the continuous-variable (CV) version of the MDI-QKD protocol can not reach/beat the PLOB bound due to the limits of the Bell detection~\cite{xu2020secure}.
In the DV community, Lucamarini \textit{et al.} extended the MDI idea to the twin-feld (TF) protocol, where the MB node performs a more advanced detection to overcome the PLOB bound~\cite{lucamarini2018overcoming}.
Unfortunately, the CV community has yet to find a way to replicate the specific interference effects that occur in TF-QKD, so that the secure transmission distance of CV-MDI-QKD is still limited (2-3 dB loss in recent experiments even with asymptotic security~\cite{tian2022experimental,hajomer2025high}).
Even so, considerable attention has been paid to CV-MDI-QKD since it is already compatible with the current telecommunication network by using the cost-effective and room-temperature homodyne detectors~\cite{fletcher2025overview}, while all the threats from the detector’s point of view are immune, e.g., calibration attack~\cite{jouguet2013preventing}, saturation attack ~\cite{qin2016quantum}, and blinding attack~\cite{qin2018homodyne}.

Unlike the DV counterpart, the asymmetric and symmetric settings of CV-MDI-QKD show unequal performance in terms of loss tolerance.
As shown in Ref.~\cite{pirandola2015high}, if Alice brought arbitrarily close to the relay (i.e., the transmittance ${\tau _A} \to 1$), the distance scaling is much improved in the asymmetric setting as Bob can be very far from the relay (the rate goes to zero for ${\tau _B} \to 0$ when Bob’s channel is affected by pure loss only). 
This feature makes CV-MDI-QKD well-suited for the typical topology of a local area network (LAN), where a user connects their device to a proxy server to communicate with remote users, as shown in Fig.~\ref{fig:MDI}(a).
%Motivated by this feature, an intuitive solution for transmission enhancement is to compensate for the transmittance of Alice’s channel, e.g., by using the optical amplifier.
Given this advantage, a straightforward approach to enhance transmission is to compensate for the channel transmittance from Alice, for instance, by employing an optical amplifier.
However, Fossier \textit{et al.}  demonstrated that the traditional amplifier amplifies the signal as well as the noise, which may reduce the signal-to-noise ratio~\cite{fossier2009improvement}.
Another solution to compensate for the loss is to use the probabilistic noiseless linear amplifier (NLA), which can in principle amplify the amplitude of a coherent state while retaining the initial level of noise~\cite{blandino2012improving,chrzanowski2014measurement}.
Unfortunately, the ideal NLA operation is unphysical, while its maximum gain show limited value at a LAN range~\cite{notarnicola2023long}.
Moreover, the probabilistic nature causes a time delay (especially for two large spatially separated nodes), as users need to alert each other of success or failure via classical communication and discard the failed instances. 

In fact, in the view of quantum information theory, the principle of NLA is to suppress the excitation loss errors by using the probabilistic two-way entanglement distillation~\cite{azuma2023quantum}.
Besides this probabilistic way, quantum error correction (QEC) is a deterministic process for error suppression that eliminates the delays associated with classical heralding signals~\cite{terhal2015quantum}.
%where a qubit is encoded into a two-dimensional subspace of a large Hilbert space composed of many physical qubits rather than directly into a single physical system. 
For large-scale quantum networks, having this determinism favorably affects communication rates.
Even though QEC can correct only a finite amount of loss errors (up only to 50\% loss error rates deterministically~\cite{stace2009thresholds,muralidharan2014ultrafast}), it is sufficient for the asymmetric CV-MDI-QKD protocol as the users in the LAN are close to the relay.
When treating the CV system, we consider the bosonic QEC codes, which can efficiently use the large Hilbert space of bosonic systems and reduce the number of bosonic modes (this might be advantageous for maximizing the usage of our optical quantum channel bandwidth~\cite{li2017cat}).
However, before 2020, most bosonic QEC schemes encoded a finite-dimensional logical qubit or qudit into noisy bosonic oscillator modes~\cite{gottesman2001encoding,michael2016new,fukui2017analog,fukui2018high}, which cannot be used for CV-QKD protocols (based on infinite-dimensional oscillator modes).
In addition, several oscillator-into-oscillators encoding schemes are proposed, but they are based on Gaussian resources and hence cannot correct practically relevant Gaussian errors (including excitation losses, thermal noise, and additive Gaussian noise errors) due to the established no-go theorems~\cite{vuillot2019quantum,eisert2002distilling,niset2009no}.

Fortunately, in 2020, Noh \textit{et al.} proposed an oscillator-into-oscillators code based on non-Gaussian Gottesman-Kitaev-Preskill (GKP) states~\cite{noh2020encoding,brady2024advances}, which can correct the practically relevant Gaussian errors and has been used for several CV quantum information tasks, e.g., distributed quantum sensing~\cite{zhuang2020distributed} and quantum illumination~\cite{wu2022continuous}.
To date, the GKP state has been successfully generated using bulk-optics setups at telecommunication wavelength~\cite{konno2024logical}, but also on an on-chip platform~\cite{larsen2025integrated}.
Motivated by the above, in this paper, we propose an enhanced asymmetric CV-MDI-QKD scheme that employs the GKP oscillators-to-oscillators codes on the LAN side.
In particular, the errors caused by both loss and operation are reduced by a corrective displacement in terms of the error syndromes extracted from the encoded GKP ancilla.
This non-Gaussian method avoids any probabilistic process and, of note, keeps the deterministic nature of a typical CV-QKD system without data discard.
Starting from a discussion over the residual errors of both noisy and noiseless GKP code, we investigate the composable finite-size security of the protocol under the collective Gaussian attack, encompassing the single-layer and concatenation structure, for both fiber and more flexible free-space links.
Assuming random permutations (so that quantum de Finetti applies), this is the most powerful attack against Gaussian protocols~\cite{weedbrook2012gaussian}.
Security analysis shows its potential to be applied in a real implementation for a broader high-rate LAN over longer secure transmission.

The rest of this paper is organized as follows.
In Sec.~\ref{sec:MDI}, we briefly introduce the general framework of CV-MDI-QKD, while showing its security with and without the optical amplifier. 
In Sec.~\ref{sec:GKP}, we investigate the enhanced scheme with the GKP oscillators-to-oscillators codes, showing its residual errors using noisy and noiseless GKP states, with the single-layer and concatenation structure.
In Sec.~\ref{sec:P}, the composable finite-size security is studied under both fiber and horizontal free-space links.
Finally, Sec.~\ref{sec:con} is for conclusions.

\section{Generic model for CV-MDI} \label{sec:MDI}

\begin{figure*}
	\begin{center}
		\includegraphics[scale=0.64]{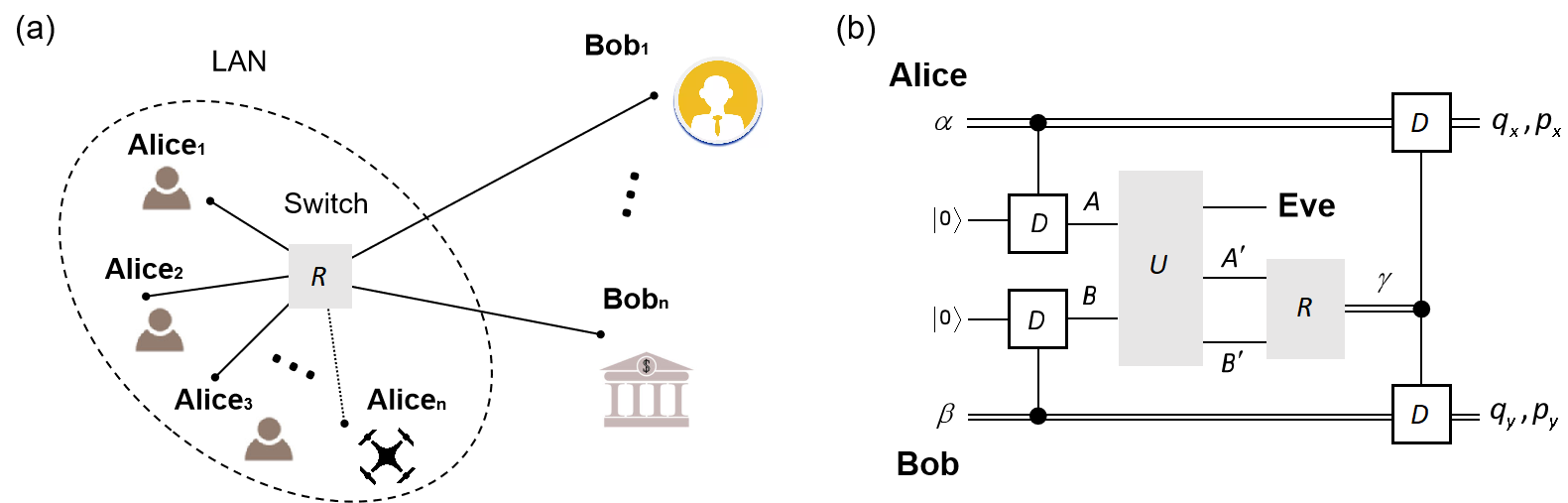}
		\caption{\label{fig:MDI} (a) Network topology using asymmetric CV MDI-QKD protocol with an untrusted relay $R$ (using active optical switch).
        LAN: local area network.
        (b) The PM scheme of the CV-MDI-QKD protocol.
Time flows from left to right, while the single lines and double lines represent the bosonic modes and classical variables, respectively.
$ \left| {\rm{0}} \right\rangle $ denotes the vacuum state.
The block $D$ is the displacement operator.
The block $U$ is Gaussian unitary involving two links.
The relay $R$ performs some (in principle unknown) physical transformation to output $ \gamma $ and gives to the eavesdropper Eve quantum side information.
The raw keys of Alice and Bob are $ \left( {{q_x},{p_x}} \right) $ and $ \left( {{q_y},{p_y}} \right)$, respectively.
		}
	\end{center}
\end{figure*}

As shown in Fig.~\ref{fig:MDI}(b), the details of Gaussian-modulated CV-MDI-QKD protocol in prepare-measure (PM) scheme are as follows.

\textit{Step 1: state preparation.}
Alice and Bob locally prepare coherent states $ \left| \alpha  \right\rangle $ and $ \left| \beta  \right\rangle $ with complex amplitudes $\alpha {\rm{ = }}{{\left( {{q_A} + i{p_A}} \right)} \mathord{\left/
 {\vphantom {{\left( {{q_A} + i{p_A}} \right)} {\rm{2}}}} \right.
 \kern-\nulldelimiterspace} {\rm{2}}}$ and $\beta {\rm{ = }}{{\left( {{q_B} + i{p_B}} \right)} \mathord{\left/
 {\vphantom {{\left( {{q_B} + i{p_B}} \right)} {\rm{2}}}} \right.
 \kern-\nulldelimiterspace} {\rm{2}}}$, respectively.
In detail, the real vectorial variables $ \left( {{q_A},{p_A}} \right) $ and $ \left( {{q_B},{p_B}} \right)$ are drawn i.i.d. from zero-mean, circular symmetric, Gaussian distributions with variances $\sigma _A^{\rm{2}}$ and $\sigma _B^{\rm{2}}$ respectively.
Then, Alice and Bob send their Gaussian-modulated coherent states to the relay through quantum channels that could be, e.g., optical fibres or free-space links.

\textit{Step 2: operations of the relay.}
At the relay, Charlie uses a CV Bell detection to perform a joint measurement on the arriving coherent states.
For each pair of coherent states, Charlie publicly announces the outcome $\gamma {\rm{ = }}{{\left( {{q_R} + i{p_R}} \right)} \mathord{\left/
 {\vphantom {{\left( {{q_R} + i{p_R}} \right)} 2}} \right.
 \kern-\nulldelimiterspace} 2}$ to Alice and Bob for conditional displacements.
The details of the outcome complex value are given by
\begin{equation}
{q_R} = \sqrt {\frac{{{\tau _B}}}{2}} {q_B} - \sqrt {\frac{{{\tau _A}}}{2}} {q_A} + {q_Z},
\end{equation}
\begin{equation}
{p_R} = \sqrt {\frac{{{\tau _B}}}{2}} {p_B} + \sqrt {\frac{{{\tau _A}}}{2}} {p_A} + {p_Z},
\end{equation}
where $q_Z$ and $p_Z$ are the noise variables,
$\tau_A$ and $\tau_B$ are the transmittance of Alice's and Bob's channel, respectively.

\textit{Step 3: parameter estimation (PE).}
Alice and Bob randomly select a portion of their real vectorial variables for channel parameters estimations (e.g., transmittance and excess noise).
For example, one can follow the method in Ref.~\cite{papanastasiou2017finite}. 
Then, Alice and Bob can compute a secret key rate with an overestimated Holevo information by using the worst-case estimators.
The details of the calculation can be found in Ref.~\cite{papanastasiou2023composable}.
If calculations show that no secret keys can be generated, they abort the protocol. 
Otherwise, they proceed.

\textit{Step 4: conditional displacements.}
In terms of $\gamma$, Alice and Bob perform conditional displacements on their real vectorial variables to get the raw keys $ \left( {{q_x},{p_x}} \right) $ and $ \left( {{q_y},{p_y}} \right)$ by using the following relations%~\cite{lupo2018parameter}
\begin{equation}\label{e:q_x}
{q_x} = {q_A} - \frac{{\left\langle {{q_A}{q_R}} \right\rangle }}{{\left\langle {q_R^2} \right\rangle }}{q_R},
\end{equation}
\begin{equation}
{p_x} = {p_A} - \frac{{\left\langle {{p_A}{p_R}} \right\rangle }}{{\left\langle {p_R^2} \right\rangle }}{p_R},
\end{equation}
\begin{equation}
{q_y} = {q_B} - \frac{{\left\langle {{q_B}{q_R}} \right\rangle }}{{\left\langle {q_R^2} \right\rangle }}{q_R},
\end{equation}
\begin{equation}\label{e:p_x}
{p_x} = {p_B} - \frac{{\left\langle {{p_B}{p_R}} \right\rangle }}{{\left\langle {p_R^2} \right\rangle }}{p_R}.
\end{equation}
The above displacement is the optimal option, which has the minimal correlation between the raw keys and the relay outcomes because one has $\left\langle {{q_x}{q_R}} \right\rangle  = \left\langle {{q_y}{q_R}} \right\rangle  = 0$ and $\left\langle {{p_x}{p_R}} \right\rangle  = \left\langle {{p_y}{p_R}} \right\rangle  = 0$.
In other words, Eve should know as less as possible about the raw keys by knowing $\gamma$.
Note that the correlation terms in Eq.~\eqref{e:q_x} to Eq.~\eqref{e:p_x} i.e., $ {\left\langle {{q_A}{q_R}} \right\rangle } $, ${\left\langle {{p_A}{q_R}} \right\rangle }$, ${\left\langle {{q_B}{q_R}} \right\rangle }$, and ${\left\langle {{p_B}{q_R}} \right\rangle }$ can be estimated by Alice or Bob locally, without the need of public communication.

\textit{Step 5: classical postprocessing.}
The raw keys are postprocessed for error correction (EC) and privacy amplification (PA).
The details of postprocessing can be found in Ref.~\cite{papanastasiou2023composable}.

\begin{table}[h]
	\centering
	\caption{Protocol parameters.}
	\label{table_P}
	\begin{tabular}{p{4cm}<{\centering}p{2cm}<{\centering}p{2cm}<{\centering}}
	\hline
	\hline
	Parameter & Symbol & Value \\
	\hline
    Signal wavelength & $\lambda$ & 1550 nm \\
         Modulation variance & $\sigma _A^{\rm{2}}, \sigma _B^{\rm{2}}$ & 20 \\
	Reconciliation efficiencye & $ \beta_0 $ & 1 \\
	SMF attention coefficient & $ \alpha_0 $ & 0.2 dB/km \\
         Thermal noise photon mean & ${\bar n}$ & 0  \\
	\hline
	\hline
	\end{tabular}
\end{table}

\begin{figure}
\centerline{\includegraphics[width=3.4in]{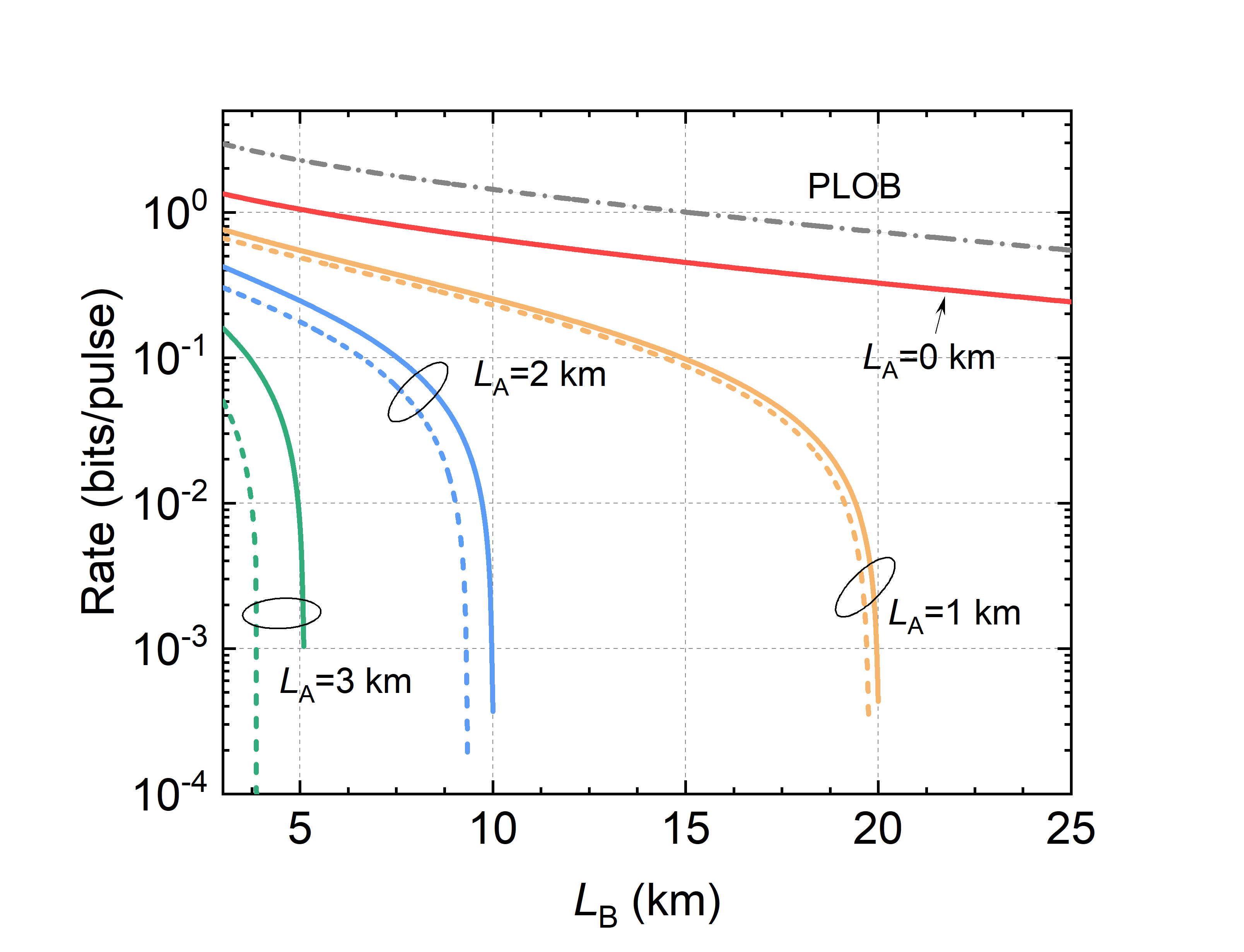}}
	\caption{\label{fig:asy} Asymptotic key rate versus fiber distance of Bob-to-relay with reverse reconciliation.
The solid lines and dashed lines represent the original configuration and the configuration with the pre-amplification operator $ {\cal A}_{1/\tau_A }^{{\rm{Pre}}} $, respectively.
The dash-dotted line denotes the PLOB bound when $L_A=0$ km.
}
\end{figure}

Fig.~\ref{fig:asy} shows the asymptotic key rate ${R_{{\rm{asy}}}}$ of the asymmetric CV-MDI-QKD protocol under the phase-insensitive collective Gaussian attack.
The protocol parameters are shown in Table~\ref{table_P}, while we follow the method in Ref.~\cite{papanastasiou2023composable} for the security analysis.
Here, the quantum channel is set to a pure-loss channel, which is connected by the standard single-mode fiber (SMF) with an attenuation coefficient of 0.2 dB/km.
Therefore, the transmittance is given by $\tau  = {10^{ - 0.2L/10}}$ with the distance $L$.
In addition, we set the detector efficiency to unit, which has reached 97.2\% in a recent experiment~\cite{tian2022experimental}.
Moreover, we also shown the case with a pre-amplification operator $ {\cal A}_{1/\tau_A }^{{\rm{Pre}}} $ on the Alice side for comparison, whose mode transform is $ \hat a \to \sqrt {1/{\tau _A}} \hat a + \sqrt {1/{\tau _A} - 1} \hat e_{1/{\tau _A}}^\dag $, where $ {{\hat e}_{1/{\tau _A}}}$ is in a vacuum state.
Then, the overall transmittance between Alice and the relay becomes $ {\tau _A} \to 1 $ (see Appendix~\ref{APP:1} for details).
We find that even $ {\tau _A} $ recovers to unit, the pre-amplification operator decreases the maximum secure transmission distance of Bob-to-relay because of the added noise from the operation. 
If no noise is added (i.e., $L_A=0$ km), the maximum transmission distance can reach 852 km.
In fact, this gap is not strange since the CV protocols appear to be fragile to noise~\cite{pirandola2020advances}.
Therefore, we discuss the tools for noise suppression in the following.
Note that we perform the amplification before the channel instead of after it since the pre-amplification causes less noise~\cite{noh2018quantum}.

\section{Tools for deterministic error suppression}  \label{sec:GKP}

%上一章加入放大器之后，信道已经变为了AWGN，可以使用GKP了

In this section, we first introduce the GKP oscillators-to-oscillators code, and then show the residual errors of tha data mode after the error correction.
In fact, when Alice uses a pre-amplification to compensate for all losses, the modulated coherent state can be regarded as passing through an AWGN channel with random displacement~\cite{noh2018quantum}, whose variance is $ {\sigma ^2} $.
In this case, the GKP code can be used to handle the excitation loss errors because the loss errors have been converted into the additive noise errors.
In detail, the overall channel after the pre-amplification becomes 
\begin{equation}\label{e:7}
{{\cal L}_{{\tau _A},{{\bar n}}}}^\circ A_{1/{\tau _A}}^{{\rm{Pre}}} = {\Phi _{{{\bar n}} + 1 - {\tau _A}}},
\end{equation}
where the thermal-loss channel $ {{\cal L}_{\tau _A ,{{\bar n}}}}$ is described by the beam-splitter transform $\hat a \to \sqrt {{\tau _A}} \hat a + \sqrt {1 - {\tau _A}} {\hat e_L}$, where the environment mode ${\hat e_L}$ has the mean photon number $\left\langle {\hat e_L^\dag {{\hat e}_L}} \right\rangle  = {\bar n}/\left( {1 - {\tau _A}} \right)$.
Therefore, the variance of the AWGN channel has ${\sigma ^2} = {{\bar n}} + 1 - {\tau _A} $.
In what follows, we consider the pure-loss channel (i.e., ${\bar n}=0$) so that the variance is $ {\sigma ^2} = 1 - \tau_A  $.
%Moreover, we assume the self-supported and identically noise model, so the signal and the GKP ancilla have the same additive noise variance $ {\sigma ^2}$.

\begin{figure*}
	\begin{center}
		\includegraphics[scale=0.58]{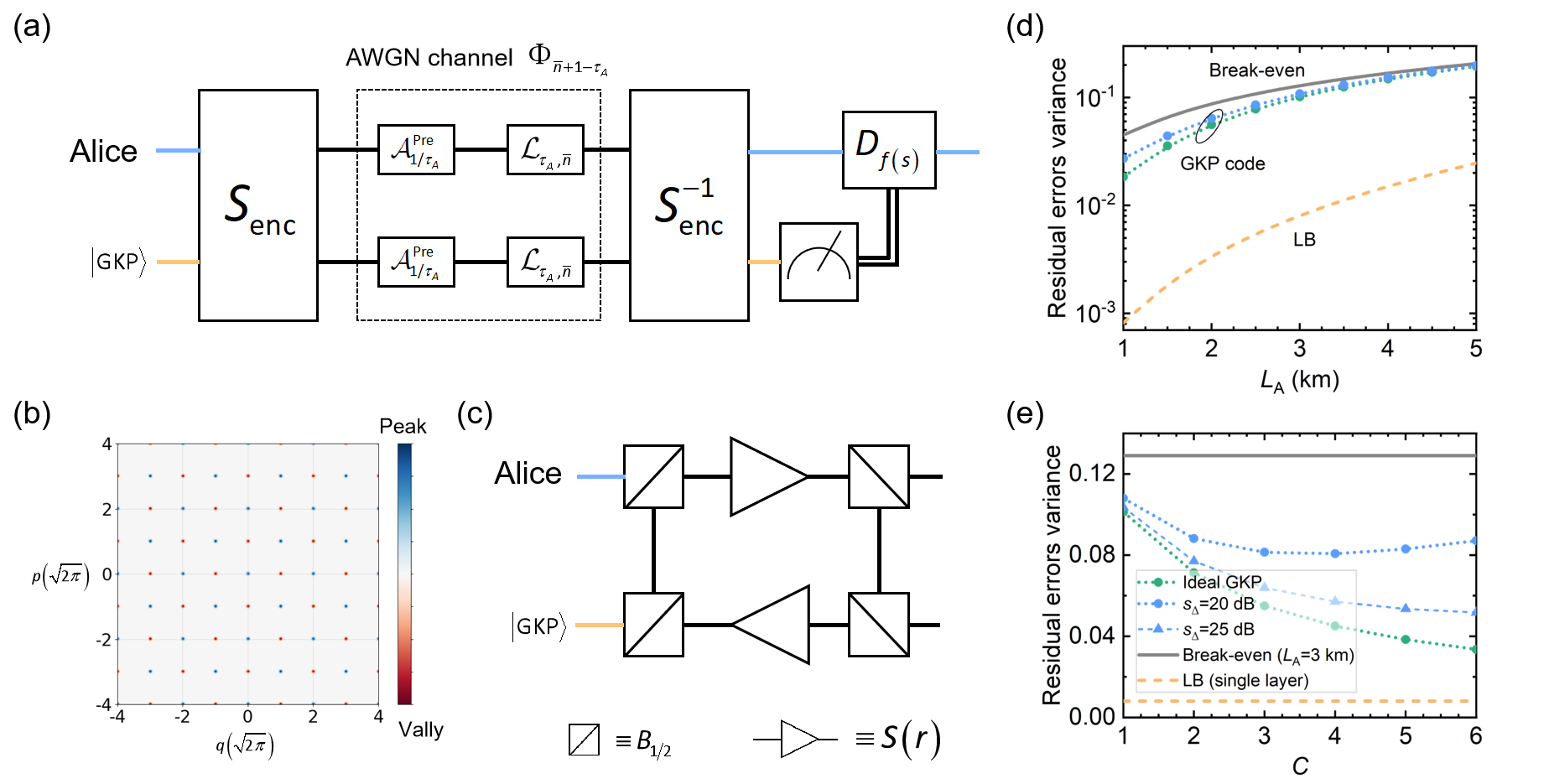}
		\caption{\label{fig:GKP}(a) Schematic of the protocol with GKP error correction on the Alice side.
        The blue lines and orange lines represent the data mode and ancilla mode, respectively.
        The GKP codes encode the data mode by a Gaussian unitary $S_{\rm enc}$.
        The overall AWGN channel $ {\Phi _{{{\bar n}_B} + 1 - {\tau _A}}} $ consists of amplifier $ {\cal A}_{1/\tau_A }^{{\rm{Pre}}}$ and thermal-loss channel $  {{\cal L}_{\tau _A ,{{\bar n}_B}}} $.
        The error syndromes $s$ are extracted from stabilizer measurements on the ancilla mode and inform the data mode for a corrective displacement $D_{f(s)}$.
        (b) The Wigner function of the ideal square GKP state.
        (c) Architecture of the TMS operation.
        ${B_{1/2}}$ denotes the two-mode beamsplitter with transmittance $1/2$.
        $S\left( r \right) $ denotes the one-mode squeezer with the squeezing parameter $r$.
        (d) Residual errors variance versus the Alice-to-relay distance with optimal $r$.
        The blue and green lines denotes the case of ideal GKP state and the approximate GKP state with a 20 dB squeezing, respectively.
        LB: lower bound.
        (e) Residual errors variance versus the connection layers number when the Alice-to-relay distance is $L_A=$3 km.
		}
	\end{center}
\end{figure*}

\subsection{Error correction with the GKP ancilla}

Here, we consider the GKP oscillators-to-oscillators codes for deterministic error suppression on the LAN side, as shown in Fig.~\ref{fig:GKP}(a).
In detail, the GKP codes encode a single data mode into two modes by entangling the data (by a Gaussian unitary) with an ancilla mode that is prepared in some non-Gaussian GKP lattice states.
As described in Ref.~\cite{gottesman2001encoding,duivenvoorden2017single}, the GKP lattice state is defined as the unique (up to an overall phase) simultaneous eigenstate of the two commuting stabilizers with unit eigenvalues.
In general, the stabilizer group of a single-mode GKP state can be expressed as
\begin{equation}
 {S_{\rm{1}}} \equiv \exp \left( {i\lambda _1^{\rm T}\Omega \hat r} \right),
\end{equation}
\begin{equation}
{S_{\rm{2}}} \equiv \exp \left( {i\lambda _2^{\rm T}\Omega \hat r} \right),  
\end{equation}
where $ {\hat r^{\rm{T}}} = \left( {{{\hat q}},{{\hat p}}} \right) $ is the vector of canonical operators,
$\Omega = (0, 1; -1, 0)$ is the single-mode symplectic form,
and the lattice vectors ($\lambda _1$, $\lambda _2$) determine the type of GKP state.
For example, the classic square (canonical) GKP state ${\left| {{\rm{GKP}}} \right\rangle _{\square}}$ has a set of vectors
\begin{equation}
\lambda _1^{\square} = \ell {\left( {0,1} \right)^{\rm{T}}},
\end{equation}
\begin{equation}
\lambda _2^{\square} = \ell {\left( {1,0} \right)^{\rm T}},
\end{equation}
where $ \ell  \equiv \sqrt {2\pi } $, such that the stabilizers are $S_1^{\square} \equiv \exp \left( {-i\sqrt {2\pi } \hat q} \right)$ and $S_2^{\square} \equiv \exp \left( {i\sqrt {2\pi } \hat p} \right)$.
Then, the square GKP state should follow $S_k^{\square}{\left| {{\rm{GKP}}} \right\rangle _{\square}}{\rm{ = }}{\left| {{\rm{GKP}}} \right\rangle _{\square}}\forall k \in \left\{ {1,2} \right\}$.
In terms of the basis states for $\hat q$ and $\hat p$, 
the ideal square GKP state is given by 
\begin{align}
\left| {{\rm{GKP}}} \right\rangle_{\square}  \propto \sum\limits_{n \in Z} {\left| { q = \sqrt {2\pi } n} \right\rangle }  \propto \sum\limits_{n \in Z} {\left| { p = \sqrt {2\pi } n} \right\rangle },
\end{align}
where $q$ and $p$ are the canonical variables.
The Wigner function of an ideal square GKP state is shown in Fig.~\ref{fig:GKP}(b).
This state can be considered as the superposition of an infinite number of position or momentum eigenstates along a grid.
In other words, this quantum state enables the joint measurement of displacements on both quadratures up to a module $\sqrt{2\pi} $ ambiguity, so that the joint
estimation of both quadrature displacements is possible (without violating the uncertainty principle)~\cite{noh2020encoding}.
In this work, we select the square GKP state as the ancilla mode, which has recently been realized in propagating light at telecommunication wavelength~\cite{konno2024logical}.
Note that other forms of GKP states with different lattice vectors, such as the hexagonal or rectangular GKP states, may also be used for practical tasks.
In fact, the lattice vectors of these GKP state can be decompose into that of square GKP state with a symplectic transformation $\Lambda  \in {\rm{Sq}}\left( {2,\mathbb{R}
} \right)$ i.e., $\left( {\Lambda \lambda _1^{\square},\Lambda \lambda _2^{\square}} \right)$~\cite{brady2024advances}.

Once the GKP state is determined, the GKP codes encode the data mode by a Gaussian unitary, which is described by the symplectic transform $ {S_{{\rm{enc}}}} $ in Fig.~\ref{fig:GKP}(a).
Like the GKP state, there are multiple choices for the symplectic transform, while our work selects the two-mode-squeezing (TMS) operation, which is shown in Fig.~\ref{fig:GKP}(c).
This selection is because Ref.~\cite{wu2023optimal} has proven that an arbitrary GKP oscillators-to-oscillators code can be reduced to a generalized GKP-TMS code.
As shown in Fig.~\ref{fig:GKP}(c), the TMS operation can be decomposed into two-mode beamsplitters and one-mode squeezers, which can be expressed as
\begin{equation}
{S_{\rm TMS}} = {B_{1/2}} \cdot \left( {S\left( r \right) \oplus S\left( { - r} \right)} \right) \cdot B_{1/2}^{\rm T},
\end{equation}
where ${B_{1/2}}$ is the symplectic representation of the two-mode beamsplitter with transmittance $1/2$,
and $S\left( r \right): = {\rm{diag}}\left( {{e^{ - r}},{e^r}} \right)$ denotes the one-mode squeezer with the squeezing parameter $r$.
After the TMS operation, the quadrature of data and ancilla mode has becomes $\hat x' \to {S_{\rm TMS}}\hat x$, where $\hat x = {\left( {{{\hat q}_A},{{\hat p}_A},{{\hat q}},{{\hat p}}} \right)^{\rm{T}}}$ and $\hat x' = {\left( {{{\hat q'}_A},{{\hat p'}_A},{{\hat q'}},{{\hat p'}}} \right)^{\rm{T}}}$ demotes the quadratures before and after the TMS operation, respectively.
Then, the encoded modes pass through the AWGN channel with the additive Gaussian noise error $ \xi {\rm{ = }}{\left( {\xi _q^d,\xi _p^d,\xi _q^a,\xi _p^a} \right)^{\rm{T}}} $, thus the arriving modes have the quadratures $\hat x'' \to \hat x' + \xi $.
Here we assume the self-supported and identically noise model so that the covariance matrix (CM) of $ \xi $ is ${V_\xi } = {\sigma ^2}{{\rm I}_4}$.

On the decoding side, the inverse transformation $S_{{\rm{enc}}}^{ - 1}$, or equivalently, $S_{\rm TMS}^{ - 1}$ is applied to disentangle the data and the ancilla.
Then, the quadrature operator is transformed into $\hat x''' \to S_{\rm TMS}^{ - 1}\hat x'' = \hat x + z$, where $z$ is the reshaped quadrature noise vector, which is given by
\begin{align}\label{e_z_GKP}
z = \left( {\begin{array}{*{20}{c}}
{\cosh \left( r \right)\xi _q^d - \sinh \left( r \right)\xi _q^a}\\
{\cosh \left( r \right)\xi _p^d - \sinh \left( r \right)\xi _p^a}\\
{\cosh \left( r \right)\xi _q^a - \sinh \left( r \right)\xi _q^d}\\
{\cosh \left( r \right)\xi _p^a - \sinh \left( r \right)\xi _p^d}
\end{array}} \right) \equiv \left( {\begin{array}{*{20}{c}}
{z_q^d}\\
{z_p^d}\\
{z_q^a}\\
{z_p^a}
\end{array}} \right),
\end{align}
where the data and ancilla noise are ${x_d} = (z_q^d\; z_p^d)$ and ${x_a} = (z_q^a\; z_p^a)$, respectively.
We find that the TMS operations correlates the noises of the data and ancilla, thus allowing for error correction via measurements on the ancilla.
In detail, one performs homodyne measurement on the ancilla system and then module $\sqrt{2\pi}$ to extract information about the data noise, which is encoded in an error syndrome~\cite{brady2024advances}
\begin{equation}\label{e:es}
s = {R_{\sqrt {2\pi } }}\left( {M_{\square}^{\rm{T}}\Omega {x_a}} \right){\mkern 1mu} {\kern 1pt}  \in {\Im ^2}, 
\end{equation}
where $ {R_{\sqrt {2\pi } }}\left( x \right) \equiv x - {n^\diamondsuit }\left( x \right)\sqrt {2\pi }  $ with $ {n^\diamondsuit }\left( x \right) \equiv \arg {\min _{n \in Z}}\left| {x - n\sqrt {2\pi } } \right| $,
the interval is defined as $\Im  \equiv [ - \sqrt {{\pi  \mathord{\left/
 {\vphantom {\pi  2}} \right.
 \kern-\nulldelimiterspace} 2}} ,\sqrt {{\pi  \mathord{\left/
 {\vphantom {\pi  2}} \right.
 \kern-\nulldelimiterspace} 2}} ]$,
and ${M_{\square}} = {\ell ^{ - 1}}\left( {\begin{array}{*{20}{c}}
{\lambda _1^{\square}}&{\lambda _2^{\square}}
\end{array}} \right) = {{\rm I}_2}$ is the generator matrix of the canonical square lattice.
Mathematically, the probability density function (PDF) of the joint distribution of the data noise and the syndrome can be expressed as
\begin{equation}\label{e:pdf}
\begin{aligned}
    {\cal P}\left( {{x_d},s} \right) & =  \sum\limits_k {g\left( {V_d^{ - 1},{x_d} + V_d^{ - 1}{V_{da}}\left( {s - k\sqrt {2\pi } } \right)} \right) \times } \\
    &  g\left( {V_{d\left| a \right.}^{ - 1},s - k\sqrt {2\pi } } \right),
\end{aligned}
\end{equation}
%\begin{widetext}
%\end{widetext}
where $g\left( {\Sigma ,x} \right)$ is the PDF of the 2-dimensional multivariate Gaussian distribution~\cite{weedbrook2012gaussian}, and the matrices above are defined implicitly via
\begin{equation}
{\left( {\begin{array}{*{20}{c}}
{{V_d}}&{{V_{da}}}\\
{V_{da}^{\rm{T}}}&{{V_a}}
\end{array}} \right)^{ - 1}}: = \left( {{{\rm{I}}_2} \oplus \Omega } \right){V_z}\left( {{{\rm{I}}_2} \oplus {\Omega ^{\rm{T}}}} \right),
\end{equation}
where ${V_{d\left| a \right.}} = {V_a} - V_{da}^{\rm{T}}V_d^{{\rm{ - 1}}}{V_{da}}$,
and $V_z$ is the CM of $z$, which follows $V_z^{ - 1} = S_{\rm TMS}^{\rm T}{V_\xi }{S_{\rm TMS}}$.
In terms of $s$, a syndrome-informed displacement operations $D_{f(s)}$ are performed on the data, ideally ridding the data of noise. 
Here $f(s)$ is an (vector) estimation function that takes the syndromes $s$ as input and provides an estimate for the error displacements on the data. 
After the error correction, the output noise on the data is ${x_{{\rm{out}}}} = {x_d} - f\left( s \right)$, whose PDF is given by 
\begin{equation}\label{e:P_out}
{\cal P}\left( {{x_{{\rm{out}}}}} \right) = \int_{{\mathbb{R}^2}} {d{x_d}} \int_{{\Im^2}} {dsP\left( {{x_d},s} \right)} \delta \left( {{x_{{\rm{out}}}} - {x_d} + f\left( s \right)} \right),
\end{equation}
where $\delta$ is the Dirac delta function.
From the distribution above, one can obtain the CM for the output error ${V_{{\rm{out}}}}{\rm{ = }}\sigma _r^{\rm{2}}{{\rm{I}}_2}$, where $\sigma _r^{\rm{2}}$ is the variance of residual errors on each quadrature.

\subsection{Residual errors}

Due to the analog nature of the errors, error correction is never perfect, and residual errors $x_{\rm out}$ will appear on the output data modes. 
The goal of error correction is to make these errors arbitrarily small so that they negligibly corrupt the data.
For this goal, first, a good estimation function is needed, as ${x_{\rm out}} \to 0$ if the function $f\left( s \right) \to {x_d}$.
So far, two typical estimators used for GKP decoding are linear estimation and minimum mean square error (MMSE) estimation, while Ref.~\cite{wu2023optimal} has proven that the MMSE estimation has higher break-even point (the point above which noise can no longer be reduced by error correction).
In this work, we select the linear estimation because while matrix multiplication and inversion are both required (only once) to derive the estimators, linear estimation involves no summations~\cite{wu2023optimal}. 
%On the contrary, MMSE estimation requires summation over two integer vectors of length 2, and the length and cost grow with the number of modes.
In the linear estimation, the estimation function is linearly related to the syndrome-i.e., $f\left( s \right) = \phi s$, where $\phi$ is some invertible matrix.
Assume that the additive noises $x_d$ and $x_a$ are small, Eq.~\eqref{e:pdf} is approximately a Gaussian distribution, so that the best choice is $\phi =  - V_d^{ - 1}{V_{da}}$~\cite{brady2024advances}.
Finally, the estimation function is given by
\begin{equation}
f\left( s \right) =  - V_d^{ - 1}{V_{da}}s = \tilde \mu \left( {\begin{array}{*{20}{c}}
0&1\\
1&0
\end{array}} \right)s,
\end{equation}
where $\tilde \mu = 2{{\cosh \left( r \right)\sinh \left( r \right)} \mathord{\left/
 {\vphantom {{\cosh \left( r \right)\sinh \left( r \right)} {\left[ {{{\cosh }^2}\left( r \right) + {{\sinh }^2}\left( r \right)} \right]}}} \right.
 \kern-\nulldelimiterspace} {\left[ {{{\cosh }^2}\left( r \right) + {{\sinh }^2}\left( r \right)} \right]}}$.

Besides a good estimation function, a high-quality GKP state is also needed for errors suppression.
Unfortunately, ideal GKP states are not normalizable and thus not physical.
Therefore, we focus on the approximate square GKP states with a finite squeezing in the following, whose wave function is given by
\begin{equation}
\begin{aligned}
    {\left| {{\rm{GKP}}} \right\rangle _{\blacksquare}} &\propto \sum\limits_{n \in Z} {{e^{ - \pi {\Delta ^2}{n^2}}}\int_{ - \infty }^\infty  {{e^{ - {{{{\left( {q - \sqrt {2\pi } n} \right)}^2}} \mathord{\left/
 {\vphantom {{{{\left( {q - \sqrt {2\pi } n} \right)}^2}} {2{\Delta ^2}}}} \right.
 \kern-\nulldelimiterspace} {2{\Delta ^2}}}}}} \left| q \right\rangle dq},  \\
    & \propto \sum\limits_{n \in Z} {{e^{ - {{{\Delta ^2}{p^2}} \mathord{\left/
 {\vphantom {{{\Delta ^2}{p^2}} 2}} \right.
 \kern-\nulldelimiterspace} 2}}}\int_{ - \infty }^\infty  {{e^{ - {{{{\left( {p - \sqrt {2\pi } n} \right)}^2}} \mathord{\left/
 {\vphantom {{{{\left( {p - \sqrt {2\pi } n} \right)}^2}} {2{\Delta ^2}}}} \right.
 \kern-\nulldelimiterspace} {2{\Delta ^2}}}}}} } \left| p \right\rangle dp,
\end{aligned}
\end{equation}
where $ \Delta $ is the variance of the additive noise associated with the finite GKP squeezing, and the GKP squeezing is then defined as ${s_\Delta } =  - 10{\log _{10}}\left( {2{\Delta ^2}} \right)$.
Physically, ${s_\Delta } $ quantifies how much an approximate GKP state is squeezed in both the position and the momentum quadrature in comparison to the vacuum noise variance 1/2.
Due to the finite squeezing, the input GKP ancilla should be modeled as an ideal GKP state with the additional noise $\xi _q^\Delta $ and $\xi _p^\Delta$ on each quadrature.
Correspondingly, the error syndrome in Eq.~\eqref{e:es} becomes
\begin{equation}
s = {R_{\sqrt {2\pi } }}\left( {M_{\square}^{\rm{T}}\Omega {x_a} + {x_\Delta }} \right),
\end{equation}
where ${x_\Delta } = {\left( {\xi _p^\Delta ,\xi _q^\Delta } \right)^{\rm{T}}}$.
Then, we can follow the same steps from Eq.~\eqref{e:pdf} to Eq.~\eqref{e:P_out} to get $\sigma _r^2$ with the noisy GKP state.
Finally, we remark that we directly use the variance $\sigma _r^2$ to quantify the performance of error correction.
Note that some works introduce the geometric mean (GM) error $\sigma _{{\rm{GM}}}^2: = \sqrt[{2M }]{{\det {V_{{\rm{out}}}}}}$ or the root-mean-square (RMS) error $\sigma _{{\rm{RMS}}}^2: = {\rm{Tr}}\left\{ {{V_{{\rm{out}}}}} \right\}/2M $ to quantify the performance of GKP codes, where $M $ is the number of data mode~\cite{brady2024advances}.
These quantities are suitable for the case with the heterogeneous and correlated noises, but this work set a self-supported and identically model so that one can easily verify $\sigma _{{\rm{GM}}}^2 = \sigma _{{\rm{RMS}}}^2 = \sigma _r^2$.

Fig.~\ref{fig:GKP}(d) shows the residual errors variance $ \sigma _r^2$ versus Alice-to-relay distance $L_A$ with optimal TMS squeezing parameter.
Here we also plot the break-even points $ \sigma _{{\rm{be}}}^2 = {\sigma ^2} $ and the lower bound (LB) of single-layer error correction for comparison, while the latter is given by~\cite{noh2018quantum}
\begin{equation}
\sigma _{{\rm{LB}}}^2 = \frac{{{\sigma ^{\rm{4}}}}}{{e{{\left( {1 - {\sigma ^{\rm{2}}}} \right)}^2}}}.  
\end{equation}
Note that the LB of the output quadrature noise is based on the quantum capacity of a general non-Gaussian additive noise channel (or equivalently, the channel’s regularized coherent information), whose derivation is shown in the supplemental material of Ref.~\cite{noh2018quantum}.
We find that GKP codes with approximate GKP states can reduce the total input noise from losses and pre-amplification when $L_A<$ 5 km.
In addition, even using an ideal GKP state, the residual error variance is one order of magnitude greater than LB.

To further suppress errors, one possible method is the concatenation, where each element in a single error-correction circuit layer can be further error-corrected by another layer of circuit.
Fig.~\ref{fig:GKP}(e) shows the residual errors variance $ \sigma _r^2$ versus the connection layers number $C$ when $ L_A=$ 3 km.
Here we consider the displacement noise continuously accumulates on the data mode, thus the residual errors with the one-by-one concatenation method is $\sigma _{{\rm{con}}}^{\rm{2}} = C\sigma _{\rm{r}}^{\rm{2}}$, where $C = {L_A}/{L_\Delta } $ with the equal inter spacing ${{L_\Delta }}$.
We find that there is an optimal layers number (i.e., $C_{\rm opt}$=4) when ${s_\Delta } $= 20 dB.
In other words, if $C>4$, the residual errors variance increases with the layers number.
Further analysis shows that this is because the low GKP squeezing as the additional noise due to finite squeezing (i.e., $\xi _q^\Delta $ and $\xi _p^\Delta$) also accumulates with the increase of the layers number. 
When GKP squeezing increases from 20 dB to 25 dB, the residual variance decreases with more connection layers.
%We find that even if only one connection layer is added, the variance of the residual error is reduced by more than half for both ideal and approximate GKP states.
%Moreover, the residual error variances of the ideal and approximate GKP states are reduced to below the LB of single-layer error correction when using 5-layer connections and 6-layer connections, respectively.
Note that one may use a more complex concatenation structure instead of the above one-by-one method for further reduction (see Ref.~\cite{zhou2022enhancing} for details), which is not the scope of this work.

\section{PERFORMANCE}\label{sec:P}

In this section, we first introduce the composable finite-size security of the enhanced scheme with GKP oscillators-to-oscillators codes.
Then, we show the composable finite-size key rate against collective attacks for both fiber and free-space scenarios.
The simulation parameters are shown in Table~\ref{table_P} and Table~\ref{tab:com}.
Here, we use the reverse reconciliation method (i.e., Alice infers Bob's data) for a higher rate as Alice approaches the relay~\cite{fletcher2025overview}.

\begin{table}[h]
	\centering
	\caption{Composable finite-size parameters.}
	\label{tab:com}
	\begin{tabular}{p{4cm}<{\centering}p{2cm}<{\centering}p{2cm}<{\centering}}
	\hline
	\hline
	Parameter & Symbol & Value \\
	\hline
    Total pulse & $ N $ & $ 10^8 $ \\
	PE signals & $ {{m_{{\rm{pe}}}}} $ & $0.1 \times N$ \\	Digitalization & $ d $ & $ 2^5 $ \\
	EC success probability & $ p_{\rm{ec}} $ & 0.9 \\
	$\varepsilon$-correctness & $ {{\varepsilon_{{\rm{cor}}}}} $ & $ 10^{-10} $ \\
	Smoothing parameter & $ {{\varepsilon_{{\rm{s}}}}} $ & $ 10^{-10} $ \\
	Hash parameter & $ {{\varepsilon_{{\rm{h}}}}} $ & $ 10^{-10} $ \\
    PE error probability & $ {{\varepsilon_{{\rm{pe}}}}} $ & $ 10^{-10} $ \\
	\hline
	\hline
	\end{tabular}
\end{table}	

\subsection{Practical composable security} 

\begin{figure}
\centerline{\includegraphics[width=3in]{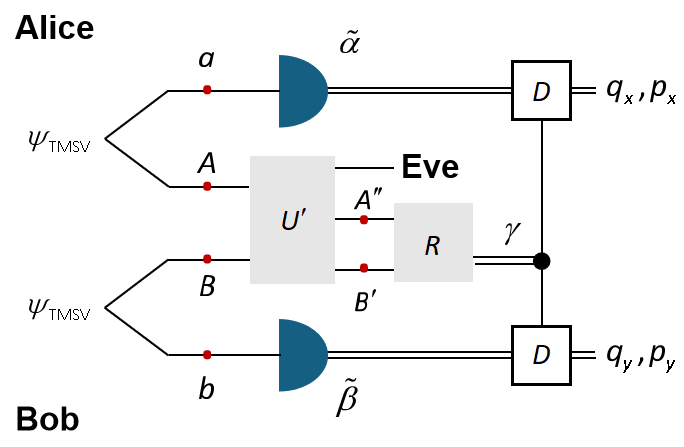}}
	\caption{\label{fig:PM} The equivalent entanglement-based scheme with error correction.
    The block $U'$ is unitary involving links, amplifier and error correction.
    The symbol ${\psi _{{\rm{TMSV}}}}$ denotes the TMSV state.
By heterodyning mode $a$ ($b$), Alice remotely prepares a coherent state $ \left| \alpha  \right\rangle $ ($ \left| \beta  \right\rangle $) on mode $A$ ($B$).
The outcomes $ \tilde \alpha $ and $ \tilde \beta $ have a linear relationship with the projected amplitude $\alpha$ and $\beta$, respectively.
}
\end{figure}

We first show the asymptotic security of the protocol under the collective attack. 
For convenience, the security analysis is based on the equivalent entanglement-based representation of the protocol (see Fig.~\ref{fig:PM}), instead of the PM version in Fig.~\ref{fig:MDI}(b).
Since Eve holds the purification of the system between Alice and Bob, the rate is completely determined by the quantum CM ${V_{ab\left| \gamma  \right.}}$ of the conditional remote state $ {\psi _{ab\left| \gamma  \right.}} $ after the Bell detection.
We remark that the property of extremality of Gaussian states implies that the knowledge of the CM is sufficient to assess the security of the protocol. %~\cite{lupo2018parameter}.
When the error correction is finished, the input global state of the Bell detection has the CM
\begin{equation}\label{e:V}
{V_{abA'B'}}{\rm{ = }}\frac{1}{2}\left( {\begin{array}{*{20}{c}}
{{V_{ab}}}&{{C_1}}&{{C_2}}\\
{C_1^T}&{{V_{A'}}}&{\bf{0}}\\
{C_2^T}&{\bf{0}}&{{V_{B'}}}
\end{array}} \right),
\end{equation}
where ${V_{ab}}{\rm{ = diag}}\left( {\sigma _A^{\rm{2}} + 1,\sigma _B^{\rm{2}} + 1} \right)$, $\textbf{0}$ is the 2 × 2 zero matrix, and the other elements are
\begin{equation}
{V_{A'}} = \left[ {{\sigma _A^{\rm{2}}} + 1 + 2\sigma _r^2} \right]{\rm{I}_2},
\end{equation}
\begin{equation}
{V_{B'}} = \left[ {{\tau _B}{\sigma _B^{\rm{2}}} + 1} \right]{\rm{I}_2},
\end{equation}
\begin{equation}
{C_1} = \left( {\begin{array}{*{20}{c}}
{\sqrt {{\sigma _A^{\rm{4}}} + 2{\sigma _A^{\rm{2}}}} {\rm{Z_2}}}\\
\textbf{0}
\end{array}} \right),
\end{equation}
\begin{equation}
{C_2} = \left( {\begin{array}{*{20}{c}}
\textbf{0}\\
{\sqrt {{\tau _B}\left( {\sigma _B^{\rm{4}} + 2{\sigma _B^{\rm{2}}}} \right)} {\rm{Z_2}}}
\end{array}} \right),
\end{equation}
where $Z_2 = {\rm{diag}}\left( {1, - 1} \right)$.
Then, we can follow the same steps in Appendix F of Ref.~\cite{pirandola2015high} for ${V_{ab\left| \gamma  \right.}}$, and finally, the asymptotic key
rate (see Appendix~\ref{APP:1} for details).

Next, we take the imperfection of data post-processing into account for a composable finite-size key rate.
The asymptotic key rate above can be calculated once Alice and Bob know the value of all the parameters entered in Eq.~\eqref{e:V}.
However, in practice, the users may not know all the parameters, but need to randomly choose and publicly disclose $m_{\rm pe}$ round of the distributed signals for PE.
Because realistic users can only use the quantum channel a finite $N$ round, this estimation is not perfect, which decreases the rate.
Here we follow the PE method in Ref.~\cite{ghalaii2022composable} to get the worst-case scenario CM $V_{ab\left| \gamma  \right.}^{{\rm{wc}}}$, which uses suitable tail bounds for the chi-squared distribution.
In particular, we do not estimate channel parameters, but use a more general way by directly estimating measurable quantities in CM.
As a result, up to an error probability $ {\varepsilon_{{\rm{pe}}}} $, Alice and Bob are able to bound the actual CM with the worst-case estimators  (see Appendix~\ref{APP:2}).
Note that the tail bounds method is suitable for the case with a low $ {\varepsilon_{{\rm{pe}}}} $, while the typical method creates divergences when $ {\varepsilon _{{\rm{pe}}}} \le {\rm{1}}{{\rm{0}}^{{\rm{ - 17}}}}$ (see Ref.~\cite{papanastasiou2017finite}).
By replacing $V_{ab\left| \gamma  \right.}^{{\rm{wc}}}$ by ${V_{ab\left| \gamma  \right.}}$, one can obtain the finite-size rate ${R_{{\rm{asy}}}}$ with the same steps in Appendix~\ref{APP:2}.

After PE, the remaining $l=N-m_{\rm pe}$ rounds are used for key generation via EC and PA.
During these steps, another related property of keys is composability, which allows specifying the security requirements for combining different cryptographic applications in a unified and systematic way~\cite{jain2022practical,zuo2026realistic}.
Mathematically, a composable security proof can be provided by incorporating proper error parameters, for each segment of the protocol, i.e., EC, PA, smoothing, and hashing.
We direct readers to Ref.~\cite{pirandola2024improved} for the details of the above steps, where the composable finite-size secret key rate is given by
\begin{equation} \label{e:com}
{R_{{\rm{com}}}} \ge \frac{{{p_{{\rm{ec}}}}\left[ {l{R_{{\rm{pe}}}} - \sqrt l {\Delta _{{\rm{aep}}}} + {{\log }_2}\left( {\varepsilon _h^2{\varepsilon _{{\rm{cor}}}}} \right)} \right]}}{N},
\end{equation}
where $ {{p_{{\rm{ec}}}}}$ is the success probability of hash comparison in EC (note that $1-p_{\rm ec}$ is known as the frame error rate),
$ {{\varepsilon _{{\rm{cor}}}}}$ bounds the probability that the sequences are different even if their hashes coincide,
the security parameter $ \varepsilon_h $ characterizes the hashing function,
and
\begin{align} 
{\Delta _{{\rm{aep}}}} \simeq 4{\log _2}\left( {\sqrt d  + 2} \right)\sqrt {{{\log }_2}\left( {2/\varepsilon _{\rm{s}}^2} \right)},
\end{align}
where $ \varepsilon_s $ is the smoothing parameter,
$d$ is the digitalization.
Finally, the total epsilon security including PE process is $\varepsilon  = {\varepsilon_{{\rm{cor}}}} + {\varepsilon_s} + {\varepsilon_h} + {p_{{\rm{ec}}}}{\varepsilon_{{\rm{pe}}}}$.
Note that Eq.~\eqref{e:com} is a lower bound to the number of secret random bits that Alice and Bob can extract, where the protocol assumes an optimal PA with $\varepsilon$-security.
It is also important to note that one can get the upper bound of composable finite-size rate in terms of the converse leftover hash bound \cite{tomamichel2011leftover}.
The upper bound means that the rate between Alice and Bob must be compressed below the value of this bound if they want to have $\varepsilon$-security assured.
When $N$ is typically large, the upper bound almost coincides with the lower bound in Eq.~\eqref{e:com}.

\subsection{Secret key rate} 

\subsubsection{Fiber scenario} 

\begin{figure*}
	\begin{center}
		\includegraphics[scale=0.85]{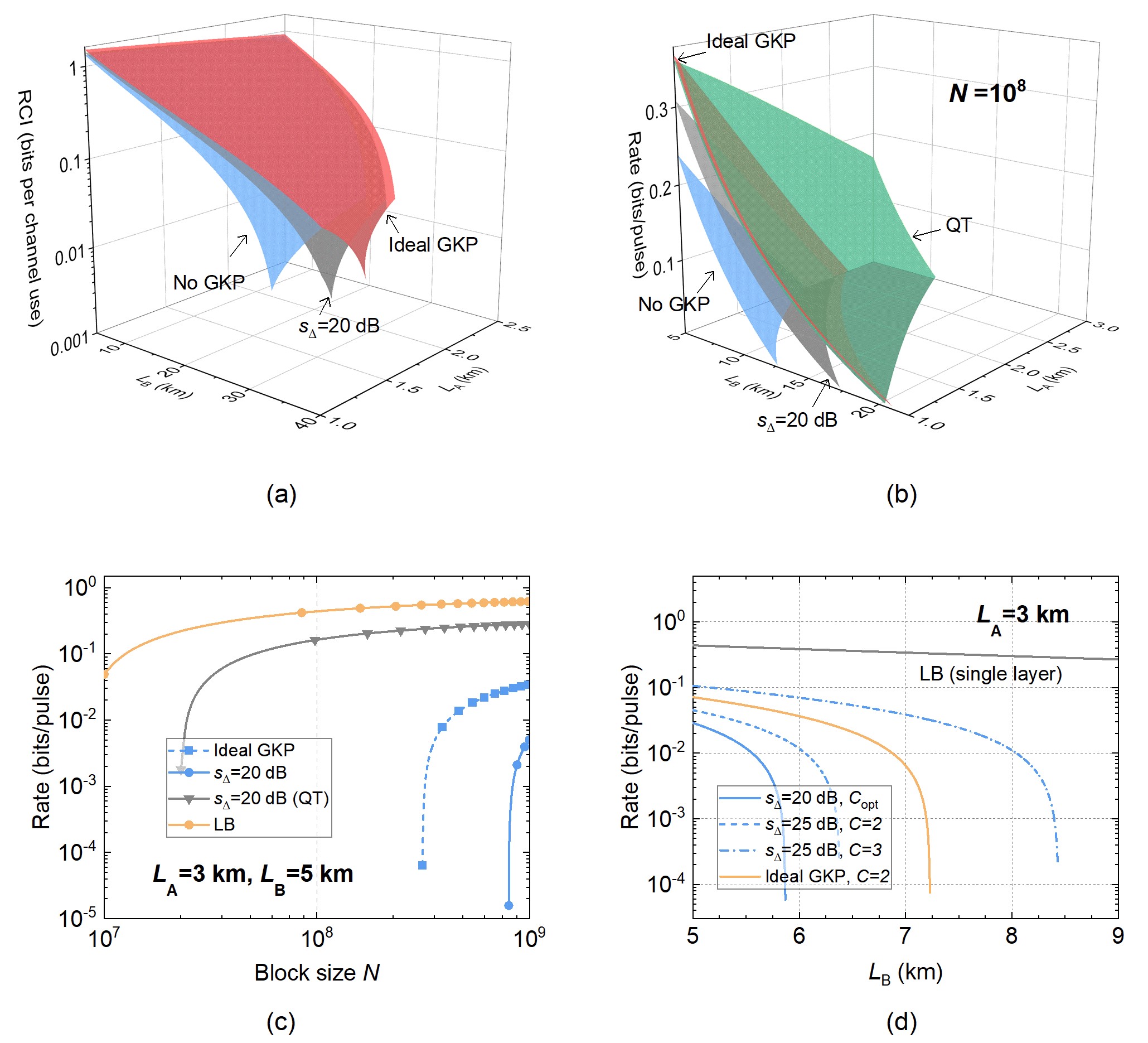}
		\caption{\label{fig:com}(a) RCI, or equivalently, the lower bound on the optimal key rate versus both Alice-to-relay and Bob-to-relay distance when using GKP code.
        (b) Composable finite-size key rate versus both Alice-to-relay and Bob-to-relay distance when using GKP code.
    %The red dashed line, blue dashed line, and green dashed line denote the secure regions of the protocol without GKP code, with the ideal GKP state, and with the QT method (20 dB TMSV squeezing), respectively.
    The green surface denotes the secure region of the protocol with the QT compaction method (20 dB TMSV squeezing).
    QT: quantum teleportation.
    (c) Composable finite-size key rate versus block size.
    The grey line denotes the typical value of the block size (i.e., $N=10^8$) used in this work.
    (d) Composable finite-size key rate versus Bob-to-relay distance with one-by-one concatenation method.
		}
	\end{center}
\end{figure*}

First, we consider the scenario where both Alice and Bob are connected to the relay using SMF.
Fig.~\ref{fig:com}(a) shows the reverse coherent information (RCI) of the bosonic system with and without GKP codes, which stands as the lower bound on the optimal key rate  $R$~\cite{pirandola2017fundamental}.
Here, we first remove the imperfection of data post-processing for the optimal rate, while the GKP code uses the optimal squeezing parameter.
For the conditional remote state $ {\psi _{ab\left| \gamma  \right.}} $, the coherent information (CI) and RCI is described by its CM ${V_{ab\left| \gamma  \right.}}$ in Eq.~\eqref{e:gama}, which can be expressed as~\cite{ghalaii2020capacity}
\begin{equation}
{I_{{\rm{CI}}}}\left( {{V_{ab\left| \gamma  \right.}}} \right){\rm{ = }}h\left( {{V_{b\left| \gamma  \right.}}} \right) - h\left( {{v_1}} \right) - h\left( {{v_2}} \right),
\end{equation}
\begin{equation}
{I_{{\rm{RCI}}}}\left( {{V_{ab\left| \gamma  \right.}}} \right){\rm{ = }}h\left( {{V_{a\left| \gamma  \right.}}} \right) - h\left( {{v_1}} \right) - h\left( {{v_2}} \right),
\end{equation}
where $ {{\rm{V}}_{a\left| \gamma  \right.}} $ and $ {{\rm{V}}_{B\left| \gamma  \right.}} $ are the reduced CM of Alice and Bob respectively, 
$ \left\{ {{v_1},{v_2}} \right\} $ is the symplextic spectrum of ${V_{ab\left| \gamma  \right.}}$, 
and  the $h$-function is defined as
\begin{equation}
h\left( {{v_k}} \right): = \frac{{{v_k} + 1}}{2}{\log _2}\frac{{{v_k} + 1}}{2} - \frac{{{v_k} - 1}}{2}{\log _2}\frac{{{v_k} - 1}}{2}.
\end{equation}
These two hence provide a lower bound on the maximum achievable rate given by
\begin{equation}
\max \left\{ {{I_{{\rm{CI}}}}\left( {{V_{ab\left| \gamma  \right.}}} \right),{I_{{\rm{RCI}}}}\left( {{V_{ab\left| \gamma  \right.}}} \right)} \right\} \le R.
\end{equation}
In our simulation, we find that the RCI is larger than the CI; thus, we plot the RCI in Fig.~\ref{fig:com}(a) to bound the optimal key rate.
We find that RCI is improved by using both ideal and noisy GKP states. 
Note that RCI is also the achievable rate for entanglement distribution using this system~\cite{ghalaii2020capacity}.

Then, we take the imperfection of data post-processing into account to show the composable finite-size rate.
Fig.~\ref{fig:com}(b) shows the composable finite-size rate $ {R_{{\rm{com}}}} $ versus the Alice-to-relay distance $L_A$ and Bob-to-relay distance $L_B$ when using a GKP code with ${s_\Delta }=$ 20 dB.
Here, we also plot the secure regions of the protocol without GKP code, with the ideal GKP state, and the scenario using quantum teleportation (QT) instead of a pre-amplification to compensate for the loss (see Appendix A of Ref.~\cite{wu2022continuous} for details).
We find that when $L_A=1$ km, the maximum $L_B$ rises from 12.7 to 17.5 km with a noisy GKP code, while further increasing to 22.5 km with the ideal GKP state.
However, even with an ideal GKP state, if an user in the LAN intends to communicate with a station 5 km away, its distance to the relay cannot exceed 2.68 km.
To extend the LAN's range, one may use the QT method to compensate for the loss, which causes less additive thermal noise but needs an additional two-mode squeezed vacuum (TMSV) source.
In detail, after QT compensation, the variance of the AWGN channel becomes ${\sigma ^{\rm{2}}}{\rm{ = }}\sqrt {{\tau _A}} {\rm{1}}{{\rm{0}}^{{{ - s_0} \mathord{\left/
 {\vphantom {{ - s} {10}}} \right.
 \kern-\nulldelimiterspace} {10}}}} + 1 - \sqrt {{\tau _A}} $ for a finite squeezing parameter $s_0$, instead of $1-\tau_A$ in Eq~\eqref{e:7}.
In Fig.~\ref{fig:com}(b), the green surface denotes the QT method with 20 dB GKP squeezing and 20 dB TMSV squeezing.
We find that the QT's secure region is lower than that of the ideal GKP state with pre-amplification when the Alice-to-relay distances are under 1.1 km, but higher after that. 
In fact, the available distance of Alice-to-relay extends to 4.6 km when the Bob-to-relay distance is 5 km.
Another method for broader LAN's range is to increase the block size $N$, as shown in Fig.~\ref{fig:com}(c).
We find that when the user in a LAN located 3 km from a relay wants to communicate with a station 5 km away, the secret key rate appears when $N$ approaches $10^9$.

Fig.~\ref{fig:com}(d) shows the composable finite-size key rate $ {R_{{\rm{com}}}}  $ versus Bob-to-relay distance $L_B$ with one-by-one concatenation method.
Here, the Alice-to-relay distance is set to 3 km, while we also plot the rate with the single-layer LB for comparison.
Note that when $L_A=$ 3 km, even with an ideal GKP state, Alice cannot communicate with users more than 5 km away from the relay by using a single-layer error correction [see Fig.~\ref{fig:com}(b)].
We find that when ${s_\Delta } $= 20 dB, the rate appears if two connection layers are added, and the maximum $L_B$ achieves 5.87 km with the optimal layer number $C_{\rm opt}=4$.
If GKP squeezing increases from 20 to 25 dB, the rate appears even if only one connection layer is added (i.e., ${{L_\Delta }}$=1.5 km), where the maximum $L_B$ is 6.38 km.
In addition, the maximum $L_B$ further increases to 8.43 km when the equal inter spacing further decreases to 1 km, which is also better than the case using two-layer error correction with the ideal GKP state.

\subsubsection{Free-space scenarios}

\begin{figure}
\centerline{\includegraphics[width=3.3in]{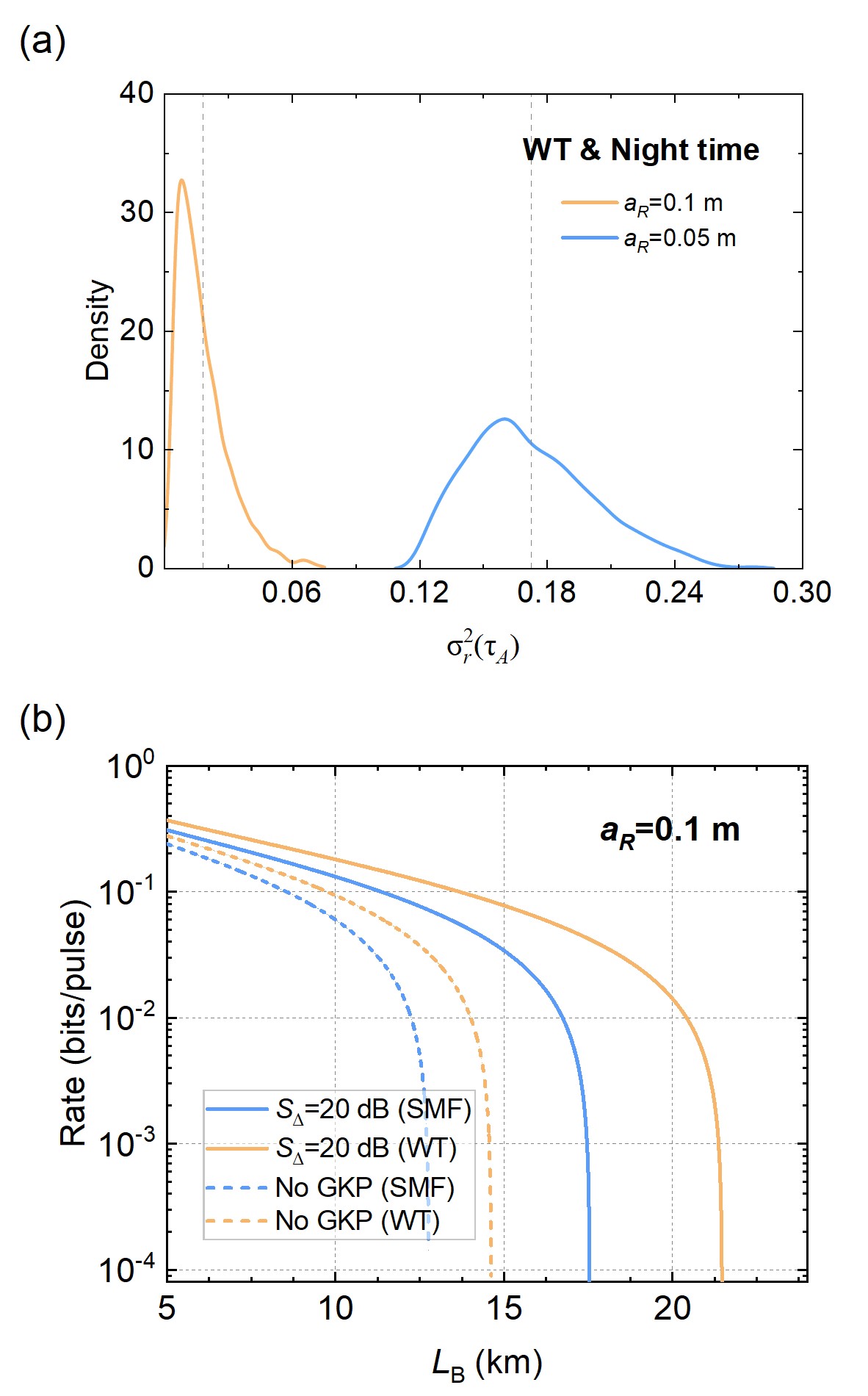}}
	\caption{\label{fig:free}(a) Probability density of ${\sigma _r^2\left( {{\tau _A}} \right)}$ over a 1 km horizontal fading channels with various receiver aperture. 
The grey lines denote the mean value of $a_R=$ 0.1 m (left) and $a_R=$ 0.05 m (right), i.e., 0.0182 and 0.1726, respectively.
    (b) Average composable finite-size key rate versus fiber distance of Bob-to-relay.
    The atmosphere structure constant is $C_n^2 = 1.28 \times {10^{ - 14}}{{\rm{m}}^{ - {2 \mathord{\left/
 {\vphantom {2 3}} \right.
 \kern-\nulldelimiterspace} 3}}}$ for night-time operation when the altitude $h_0$=30 m.
 Other parameters are the curvature ${R_0} = \infty $ (collimated Gaussian beam), the field spot size ${w_0} = 0.05 $ m, and a typical 1 $ \mu {\rm{rad}} $ pointing error.
    SMF: single-mode fiber;
    WT: week turbulence.}
\end{figure}

Next, we extend to the free-space scenario, where Alice uses a horizontal fading link to connect to the relay (Bob still uses SMF).
In this scenario, Alice can be located in a building, but also be moving aircraft or drones with more flexible links.
Unlike the fiber scenario, the transmittance along a fading channel is not fixed, but instead follows a probability distribution described by the dynamics of the environment, which poses a challenge for error correction.
Here, we assume Alice uses a dynamic GKP code, where the squeezing parameter $r$ is optimized in real time in terms of the real-time transmittance.
This assumption is possible since the frequency of signal preparation and detection (reached GHz in recent study~\cite{hajomer2024continuous}) is much higher than the fluctuation frequency of the atmospheric channel (usually 1 KHz~\cite{dequal2021feasibility}), thus, a real-time transmittance monitoring is possible (for example, by detecting the intensity of the multiplexing beacon).
With the dynamic GKP error correction, the quantum CM over fading channels has the explicit form
\begin{equation} 
V_{ab\left| \gamma  \right.}^{{\rm{fad}}} = \left( {\begin{array}{*{20}{c}}
{\Phi {{\rm{I}}_2}}&{\psi {\rm{Z_2}}}\\
{\psi {\rm{Z_2}}}&{\varphi {{\rm{I}}_2}}
\end{array}} \right),
\end{equation}
where we have set
\begin{equation}
\Phi  = \sigma _A^{\rm{2}}{\rm{ + 1}} - \left( {\sigma _A^{\rm{4}}{\rm{ + 2}}\sigma _A^{\rm{2}}} \right)\Xi ,\;{\mkern 1mu} \;{\mkern 1mu} \;{\mkern 1mu} \;{\mkern 1mu} \psi  = \sqrt {{\tau _B}} \Xi,
\end{equation}
\begin{equation}
\varphi  = \sigma _B^{\rm{2}}{\rm{ + 1}} - {\tau _B}\left( {\sigma _B^{\rm{4}}{\rm{ + 2}}\sigma _B^{\rm{2}}} \right)\Xi ,
\end{equation}
with the element
\begin{equation}
\Xi  = \int_{\rm{0}}^{{\tau _0}} {d{\tau _A}\frac{{{\cal P}\left( {{\tau _A}} \right)}}{{\sigma _A^2 + 2\sigma _r^2\left( {{\tau _A}} \right) + {\tau _B}\sigma _B^2 + 2}}},
\end{equation}
where ${{\tau _{\rm{0}}}}$ is the maximum value of the link transmittance (i.e., the transmittance when the beam centroid is perfectly aligned with the center of the receiver’s aperture),
${{\cal P}\left( {{\tau _A}} \right)}$ is the PDF of $ {\tau _A} $.
Note that here we also consider the horizontal link as a pure-loss channel, as in the above fiber case.
This assumption is feasible.
On the one hand, the short-distance horizontal link can be considered as a line-of-sight link, where Alice can employ monitoring techniques to restrict Eve, such as light detection and ranging (the same system and the corresponding optics that are being used for tracking and acquisition purposes can also be used to detect unwanted objects along the beam~\cite{ghalaii2023satellite}).
On the other hand, in CV-QKD, the bandwidth of the LO can be made very narrow by the current lasers so that the background noise can be suppressed, as only thermal noise mode-matching with the LO will survive in the output of the protocol~\cite{zuo2020atmospheric}.
As shown in Ref.~\cite{pirandola2021limits}, the night-time background noise becomes negligible with a filter $\Delta \lambda {\rm{ = 0}}{\rm{.1}}$ pm around 800 nm.

In what follows, we use the channel model in Ref.~\cite{pirandola2021limits} to get the transmittance distribution under weak turbulence (WT) and night-time operation, which also takes diffraction, extinction, and pointing errors into account.
In detail, we consider a 1 km horizontal WT link with a fixed altitude $h_0$=30 m, while the signal wavelength is changed to 800 nm to ensure good atmospheric transmission (with the extinction factor ${\alpha _{\rm{0}}} \simeq {\rm{5}} \times {\rm{1}}{{\rm{0}}^{{\rm{ - 6}}}}\;{{\rm{m}}^{ - 1}} $~\cite{vasylyev2019satellite}).
In addition, the pointing errors from jitter and imprecise tracking, which cause centroid wandering with a slow timescale, is set to typical 1 $ \mu {\rm{rad}} $ at the transmitter so that lead to a beam wandering variance $ \simeq {\left( {{\rm{1}}{{\rm{0}}^{{\rm{ - 6}}}}{L_A}} \right)^2}$.
Assume that the wandering is around an average deflection point at a distance zero from the center of the receiver’s aperture (which can be realized by means of sufficiently fast adaptive optics), the PDF for $\tau_A$ is given by
\begin{equation}\label{PDF}
{\cal P}\left( {{\tau _A}} \right) = \frac{{r_0^2}}{{\gamma_0 \sigma _{{\rm{bw}}}^2{\tau _A}}}{\left( {\ln \frac{{{\tau _{\rm{0}}}}}{{{\tau _A}}}} \right)^{\frac{2}{\gamma_0 } - 1}}\exp \left[ { - \frac{{r_0^2}}{{2\sigma _{{\rm{bw}}}^2}}{{\left( {\ln \frac{{{\tau _{\rm{0}}}}}{{{\tau _A}}}} \right)}^{\frac{2}{\gamma_0 }}}} \right],
\end{equation}
where ${\sigma _{{\rm{bw}}}^2}$ is the total centroid wanders variance (including pointing errors and turbulence),
$\gamma_0$ and $r_0$ are shape and scale (positive) parameters, respectively (see Appendix D of Ref.~\cite{pirandola2021limits} for details).
In terms of Eq.~\eqref{PDF}, we plot the PDF of ${\sigma _r^2}(\tau_A)$ over fading channels with various receiver apertures $a_R$ in Fig.~\ref{fig:free}(a).

Fig.~\ref{fig:free}(b) shows the average composable finite-size key rate versus fiber distance of Bob-to-relay when using GKP code.
For comparison, we also plot the case of connecting Alice and the relay with a SMF of the same length. 
We find that the maximum $L_B$ is longer in the free-space scenario, which approaches 20 km.
Further analysis shows that this is because the 1 km horizontal WT link has better quality than a 1 km SMF channel.
The average transmittance of the former is 0.9628, while the latter has a transmittance of 0.955.
Note that we find that Alice cannot communicate with users more than 5 km away from the relay by using a receiver aperture $a_R=0.05$ m.
In addition, we remark that Fig.~\ref{fig:free} packages the data into a package for data processing.
One may obtain better performance by clustering the data into several packages and performing security analysis for each package separately, as this can reduce the fading effect (refer to Ref.~\cite{ruppert2019fading} for details).

\section{Conclusion}\label{sec:con}

In this paper, we have proposed an enhanced scheme for the asymmetric CV-MDI-QKD protocol by using the GKP-TMS code with linear estimation.
In detail, the error from the losses and the pre-amplification are encoded in an error syndrome, which is extracted from stabilizer measurements on the ancilla mode and then reduced by a syndrome-informed displacement on the data mode.
We have investigated the residual errors of GKP codes under various GKP squeezing, showing that it can reduce the errors below the break-even point for a typical LAN range even with a noisy GKP state. 
In addition, we have also demonstrated that the residual errors can be further reduced by a simple one-by-one concatenation method, where the optimal layers number is related to the finite GKP squeezing. 
We have analyzed the composable finite-size key rates that are achieved under collective attacks, encompassing noisy and noiseless GKP states, with both SMF and horizontal links.
Our numerical results have indicated that the proposed scheme improves the maximum secure transmission range on both Alice's and Bob's side, while has better performance under a short-distance WT link with dynamic error correction.

\section*{Acknowledgments}

Z.Z. thanks Bo-Han Wu for the helpful discussion. 
S.P. acknowledges support from UKRI via the Integrated Quantum Networks Research Hub (IQN, EP/Z533208/1).
Z.Z. is supported by Quantum Science and Technology-National Science and Technology Major Project (Grant No. 2021ZD0300703).

\appendix 

\section{Asymptotic key rate}\label{APP:1}

We first derive the quantum CM ${V_{ab\left| \gamma  \right.}}$ since the rate is completely determined by it.
The reader not familiar with the formalism of bosonic systems and Gaussian states can find these concepts in Ref~\cite{weedbrook2012gaussian}.
The notations adopted here are Planck’s constant $ \hbar $ = 1 and 1/2 vacuum noise.
To derive ${V_{ab\left| \gamma  \right.}}$, we use the transformation rules for CMs under Bell-like measurements given in Ref.~\cite{spedalieri2013covariance}, which can be expressed as
\begin{equation}\label{e:gama}
{V_{ab\left| \gamma  \right.}} = {V_{ab}} - \frac{1}{{2\det \Theta }}\sum\limits_{i,j = 1}^2 {{C_i}} \left( {X_i^T\Theta {X_j}} \right)C_j^T,
\end{equation}
where 
\begin{equation}
{X_1}: = \left( {\begin{array}{*{20}{c}}
0&1\\
1&0
\end{array}} \right),{X_2}: = \left( {\begin{array}{*{20}{c}}
0&1\\
{ - 1}&0
\end{array}} \right),
\end{equation}
and the theta matrix $\Theta=\rm diag(\theta/2,\theta/2)$.
When using the GKP code, $\theta$ can be derived from Eq.~\eqref{e:V}, which has the following explicit form
\begin{equation}
\theta  = {{\left( {\sigma _A^2 + 2\sigma _r^2 + {\tau _B}\sigma _B^2 + 2} \right)} \mathord{\left/
 {\vphantom {{\left( {\sigma _A^2 + 2\sigma _r^2 + {\tau _B}\sigma _B^2 + 2} \right)} 2}} \right.
 \kern-\nulldelimiterspace} 2}.
 \end{equation}
When only the pre-amplification is used, $\theta$ becomes
\begin{equation}
\theta  = {{\left( {\sigma _A^2 - 2{\tau _A} + {\tau _B}\sigma _B^2 + 4} \right)} \mathord{\left/
 {\vphantom {{\left( {\sigma _A^2 - 2{\tau _A} + {\tau _B}\sigma _B^2 + 4} \right)} 2}} \right.
 \kern-\nulldelimiterspace} 2}.
 \end{equation}
Note that for any outcome $\gamma$, the conditional remote state $ {\psi _{ab\left| \gamma  \right.}} $ has always the same CM ${V_{ab\left| \gamma  \right.}}$, while its mean value varies with $\gamma$.

In the asymptotic regime ($N \gg 1$, where $N$ is the total number of received pulse), the secret key rate follows the Csiszar-Korner theorem given by \cite{csiszar2003broadcast}
 \begin{equation}
   R_{\rm asy}=\beta_0 I-\chi,  
 \end{equation}
where $\beta_0$ represents the reconciliation efficiency, $I$ represents the mutual information between Alice and Bob, and $\chi$ is the Holevo bound, representing the maximum information Eve has stolen.
With the reverse reconciliation method, the mutual information is given by
\begin{equation}
I = \frac{1}{2}{\log _2}\left( {\frac{{1 + \det {{\rm{V}}_{b\left| \gamma  \right.}} + {\rm{Tr}}{{\rm{V}}_{b\left| \gamma  \right.}}}}{{1 + \det {{\rm{V}}_{b\left| {\gamma \tilde \alpha } \right.}} + {\rm{Tr}}{{\rm{V}}_{b\left| {\gamma \tilde \alpha } \right.}}}}} \right),
\end{equation} 
where $ {{\rm{V}}_{\left. b \right|\gamma \tilde \alpha }} $ is Bob's CM conditioned to mode $a$'s outcome ${\tilde \alpha }$, which is given by
\begin{equation}
{{\rm{V}}_{\left. b \right|\gamma \tilde \alpha }} = {{\rm{V}}_{b\left| \gamma  \right.}} - {C_{ab\left| \gamma  \right.}}{({{\rm{V}}_{a\left| \gamma  \right.}} + {\rm{I}})^{ - 1}}C_{ab\left| \gamma  \right.}^{\rm{T}},
\end{equation} 
where $ C_{ab} $ comes from ${V_{ab\left| \gamma  \right.}}$ (the correlation between mode $a$ and mode $b$).
Next, $\chi$ is given by
\begin{equation}\label{echibE}
\chi  = S\left( {{\rho _{ab\left| \gamma  \right.}}} \right) - S\left( {{\rho _{\left. b \right|\gamma \tilde \alpha }}} \right),
\end{equation}
where $S\left(  \cdot  \right)$ denotes the Von Neumann entropy.
The first entropy term can be computed from the symplextic spectrum $ \left\{ {{v_1},{v_2}} \right\} $ of ${V_{ab\left| \gamma  \right.}}$, thus we have
\begin{equation}
S\left( {{\rho _{ab\left| \gamma  \right.}}} \right) = h({v_1}) + h({v_2}).
\end{equation}
Then, the second entropy term in Eq. (\ref{echibE}) can be computed from the symplextic spectrum $ {v_3} $ of $ {{\rm{V}}_{\left. b \right|\gamma \tilde \alpha }} $, thus we have
\begin{equation}
S\left( {{\rho _{\left. b \right|\gamma \tilde \alpha }}} \right) = h({v_3}).
\end{equation}

\section{The worst-case scenario CM}\label{APP:2}

First, the symmetric matrix ${V_{ab\left| \gamma  \right.}}$ can also be represented by the following form
\begin{equation}\label{e:Vab|xy_form}
{V_{ab\left| \gamma  \right.}}{\rm{ = }}\left( {\begin{array}{*{20}{c}}
{\left\langle {q_a^2} \right\rangle }&0&{\left\langle {{q_a}{q_b}} \right\rangle }&0\\
{...}&{\left\langle {p_a^2} \right\rangle }&0&{\left\langle {{p_a}{p_b}} \right\rangle }\\
{...}&{...}&{\left\langle {q_b^2} \right\rangle }&0\\
{...}&{...}&{...}&{\left\langle {p_b^2} \right\rangle }
\end{array}} \right).
\end{equation}
For the PE process, we follows the method in Ref.~\cite{ghalaii2022composable}, which uses the tail bounds
\begin{equation}\label{e44}
\Pr \left[ {Q \ge {m_{{\rm{pe}}}} + 2\sqrt {{m_{{\rm{pe}}}}\kappa }  + 2\kappa } \right] \le \exp \left( { - \kappa } \right),
\end{equation}
\begin{equation}\label{e45}
\Pr \left[ {Q \le {m_{{\rm{pe}}}} - 2\sqrt {{m_{{\rm{pe}}}}\kappa } } \right] \le \exp \left( { - \kappa } \right),
\end{equation}
where $Q$ denotes the random variable follows the chi-squared distribution ${\chi ^2}\left( {{m_{{\rm{pe}}}},0} \right)$,
$ \kappa $ is related to the error probability of PE, i.e., $ {{\varepsilon _{{\rm{PE}}}}} $. 
Note that, in principle, Alice and Bob can locally compute $ {\left\langle {q_a^2} \right\rangle } $, $ {\left\langle {p_a^2} \right\rangle } $, and $ {\left\langle {q_b^2} \right\rangle } $, $ {\left\langle {p_b^2} \right\rangle } $, respectively. 
Therefore, only $ {\left\langle {{q_a}{q_b}} \right\rangle } $, $ {\left\langle {{p_a}{p_b}} \right\rangle } $ in Eq. (\ref{e:Vab|xy_form}) need to be estimated by sharing part of data i.e., $m_{\rm{pe}}$ (these data can be easily acquired by Eve and thus must not contribute to the key generation).
In detail, the estimators of $ {\left\langle {{q_a}{q_b}} \right\rangle } $ and $ {\left\langle {{p_a}{p_b}} \right\rangle } $ are given by
\begin{widetext}
\begin{equation}\label{e:qaqb}
\widehat {\left\langle {{q_a}{q_{{b}}}} \right\rangle } = m_{{\rm{pe}}}^{ - 1}\sum\limits_{i = 1}^{{m_{{\rm{pe}}}}} {{{\left[ {{q_a}} \right]}_i}{{\left[ {{q_{{b}}}} \right]}_i}}  = \frac{1}{4}\left[ {{\mu _{{q_ + }}}m_{{\rm{pe}}}^{ - 1}\sum\limits_{i = 1}^{{m_{{\rm{pe}}}}} {\left[ {{{\rm{q}}_ + }} \right]_i^2 - {\mu _{{q_ - }}}m_{{\rm{pe}}}^{ - 1}\sum\limits_{i = 1}^{{m_{{\rm{pe}}}}} {\left[ {{{\rm{q}}_ - }} \right]_i^2} } } \right],
\end{equation}
\begin{equation}\label{e_pBpb1pe}
\widehat {\left\langle {{p_a}{p_{{b}}}} \right\rangle } = m_{{\rm{pe}}}^{ - 1}\sum\limits_{i = 1}^{{m_{{\rm{pe}}}}} {{{\left[ {{p_a}} \right]}_i}{{\left[ {{p_{{b}}}} \right]}_i}}  = \frac{1}{4}\left[ {{\mu _{{p_ + }}}m_{{\rm{pe}}}^{ - 1}\sum\limits_{i = 1}^{{m_{{\rm{pe}}}}} {\left[ {{{\rm{p}}_ + }} \right]_i^2 - {\mu _{{p_ - }}}m_{{\rm{pe}}}^{ - 1}\sum\limits_{i = 1}^{{m_{{\rm{pe}}}}} {\left[ {{{\rm{p}}_ - }} \right]_i^2} } } \right],
\end{equation}
\end{widetext}
with the defined standard normal variables (follow chi-square distribution)
\begin{equation}\label{e:st1}
{\left[ {{{\rm{q}}_ \pm }} \right]_i} = \left( {{{\left[ {{q_a}} \right]}_i} \pm {{\left[ {{q_{{b}}}} \right]}_i}} \right)/\sqrt {{\mu _{{q_ \pm }}}},
\end{equation}
\begin{equation}\label{e:st2}
\left[ {{{\rm{p}}_ \pm }} \right]_i^2 = \left( {{{\left[ {{p_a}} \right]}_i} \pm {{\left[ {{p_{{b}}}} \right]}_i}} \right)/\sqrt {{\mu _{{p_ \pm }}}},
\end{equation}
where $ {{\mu _{{q_ \pm }}}} $ and $ {{\mu _{{p_ \pm }}}} $ are the variance of $ {{{\left[ {{q_a}} \right]}_i} \pm {{\left[ {{q_b}} \right]}_i}} $ and $ {{{\left[ {{p_a}} \right]}_i} \pm {{\left[ {{p_b}} \right]}_i}} $ respectively, which are given by
\begin{equation}
{\mu _{{{\rm{q}}_ \pm }}} = \left\langle {q_a^2} \right\rangle  + \left\langle {q_{{b}}^2} \right\rangle  \pm 2\left\langle {{q_a}{q_{{b}}}} \right\rangle,
\end{equation}
\begin{equation}
{\mu _{{{\rm{p}}_ \pm }}} = \left\langle {p_a^2} \right\rangle  + \left\langle {p_{{b}}^2} \right\rangle  \pm 2\left\langle {{p_a}{p_{{b}}}} \right\rangle.
\end{equation}
Note that both $ {{{\left[ {{q_a}} \right]}_i} \pm {{\left[ {{q_b}} \right]}_i}} $ and $ {{{\left[ {{p_a}} \right]}_i} \pm {{\left[ {{p_b}} \right]}_i}} $ are zero-mean Gaussian variables since ${{\left[ {{q_a}} \right]}_i}$, ${{\left[ {{q_b}} \right]}_i}$, ${{\left[ {{p_a}} \right]}_i}$ and ${{\left[ {{p_b}} \right]}_i}$ are assumed to be Gaussian.
We then impose that the estimator ${\widehat {\left\langle {{q_a}{q_b}} \right\rangle }}$ is smaller than its worst-case scenario value, and the opposite for estimator $ {\widehat {\left\langle {{p_a}{p_b}} \right\rangle }}$, i.e., 
\begin{equation}\label{e:wc}
\widehat {\left\langle {{q_a}{q_b}} \right\rangle } < {\left\langle {{q_a}{q_b}} \right\rangle _{{\rm{wc}}}}, \;{\mkern 1mu} \;{\mkern 1mu} \;{\mkern 1mu} \widehat {\left\langle {{p_a}{p_b}} \right\rangle } > {\left\langle {{p_a}{p_b}} \right\rangle _{{\rm{wc}}}}.
\end{equation}
The difference between ${\widehat {\left\langle {{q_a}{q_b}} \right\rangle }}$ and $ {\widehat {\left\langle {{p_a}{p_b}} \right\rangle }}$ is because ${{{\left\langle {{p_a}{p_b}} \right\rangle }_{{\rm{wc}}}}}$ is a negative quantity so that the corresponding inequality has different direction.
In terms of the tail bounds in Eq. (\ref{e44}) and Eq. (\ref{e45}), the standard normal variables in Eq. (\ref{e:st1}) and Eq. (\ref{e:st2}) should follow
\begin{equation}
\sum\limits_{i = 1}^{{m_{{\rm{pe}}}}} {\left[ {{{\rm{q}}_ + }} \right]_i^2,} \sum\limits_{i = 1}^{{m_{{\rm{pe}}}}} {\left[ {{{\rm{p}}_ + }} \right]_i^2}  \le {m_{{\rm{pe}}}} - 2\sqrt {{m_{{\rm{pe}}}}\kappa },
\end{equation}
\begin{equation}
\sum\limits_{i = 1}^{{m_{{\rm{pe}}}}} {\left[ {{{\rm{q}}_ - }} \right]_i^2,} \sum\limits_{i = 1}^{{m_{{\rm{pe}}}}} {\left[ {{{\rm{p}}_ - }} \right]_i^2}  \ge {m_{{\rm{pe}}}} + 2\sqrt {{m_{{\rm{pe}}}}\kappa }  + 2\kappa.
\end{equation}
Combining these equations with Eq. (\ref{e:qaqb}) to Eq. (\ref{e_pBpb1pe}), up to ${O\left( {1/{m_{{\rm{pe}}}}} \right)}$, we have

\begin{equation}
{\left\langle {{q_a}{q_b}} \right\rangle _{{\rm{wc}}}} = \frac{1}{4}\left[ {\left( {{\mu _{{q_ + }}} - {\mu _{{q_ - }}}} \right) - 2\sqrt {\kappa /{m_{{\rm{pe}}}}} \left( {{\mu _{{q_ + }}} + {\mu _{{q_ - }}}} \right)} \right],
\end{equation}
\begin{equation}
{\left\langle {{p_a}{p_b}} \right\rangle _{{\rm{wc}}}} = \frac{1}{4}\left[ {\left( {{\mu _{{p_ + }}} - {\mu _{{p_ - }}}} \right) + 2\sqrt {\kappa /{m_{{\rm{pe}}}}} \left( {{\mu _{{p_ + }}} + {\mu _{{p_ - }}}} \right)} \right].
\end{equation}
To get worst-case scenario CM, in Eq. (\ref{e:Vab|xy_form}), we replace initial means $ \left\langle  \cdot  \right\rangle $ by $ {\left\langle  \cdot  \right\rangle _{{\rm{wc}}}} $, which implies
\begin{equation}
V_{ab\left| \gamma  \right.}^{{\rm{wc}}}{\rm{ = }}\left( {\begin{array}{*{20}{c}}
{\left\langle {q_a^2} \right\rangle }&0&{{{\left\langle {{q_a}{q_b}} \right\rangle }_{{\rm{wc}}}}}&0\\
{...}&{\left\langle {p_a^2} \right\rangle }&0&{{{\left\langle {{p_a}{p_b}} \right\rangle }_{{\rm{wc}}}}}\\
{...}&{...}&{\left\langle {q_b^2} \right\rangle }&0\\
{...}&{...}&{...}&{\left\langle {p_b^2} \right\rangle }
\end{array}} \right).
\end{equation}
Note that the ideas behind the PE method described above are not limited to the MDI protocol but can also be adapted to more challenging large-scale quantum network scenarios~\cite{ghalaii2022composable}.

As shown in Eq.~\eqref{e:qaqb}, the estimation of ${\widehat {\left\langle {{q_a}{q_b}} \right\rangle }}$ is fail if either ${\left[ {{{\rm{q}}_ + }} \right]_i^2}$ or ${\left[ {{{\rm{q}}_ - }} \right]_i^2}$ fails to estimate.
Similarly, if either ${\left[ {{{\rm{p}}_ + }} \right]_i^2}$ or ${\left[ {{{\rm{p}}_ - }} \right]_i^2}$ fails to estimate, then $ {\widehat {\left\langle {{p_a}{p_b}} \right\rangle }}$ fails to estimate [see Eq.~\eqref{e_pBpb1pe}].
Therefore, the probability of the inequality in Eq.~\eqref{e:wc} is
\begin{equation}
\Pr \left[ {\widehat {\left\langle {{q_a}{q_b}} \right\rangle } < {{\left\langle {{q_a}{q_b}} \right\rangle }_{{\rm{wc}}}}} \right] \le 2\exp \left( { - \kappa } \right),
\end{equation}
\begin{equation}
\Pr \left[ {\widehat {\left\langle {{p_a}{p_b}} \right\rangle } > {{\left\langle {{p_a}{p_b}} \right\rangle }_{{\rm{wc}}}}} \right] \le 2\exp \left( { - \kappa } \right).
\end{equation}
Finally, the total probability of failure is $\le 4\exp \left( { - \kappa } \right) $, which is the bound on the probability that the CM is worse than $V_{ab\left| \gamma  \right.}^{{\rm{wc}}}$ (in which case the rate would be less than the worst-case value). 
In fact, the parties can only allow this to happen with a very small probability that is less than ${\varepsilon _{{\rm{pe}}}}$, thus we have
\begin{equation}
4\exp \left( { - \kappa } \right) \le {\varepsilon _{{\rm{pe}}}},
\end{equation}
which defines $\kappa  = {4 \mathord{\left/
 {\vphantom {4 {{\varepsilon _{{\rm{pe}}}}}}} \right.
 \kern-\nulldelimiterspace} {{\varepsilon _{{\rm{pe}}}}}}$.

\bibliography{name.bib}

\end{document}